\documentclass[12pt,a4]{article}
\usepackage[a4paper]{geometry}
\usepackage[utf8]{inputenc}
\usepackage{fontenc}
\usepackage{setspace}    
\usepackage{amsmath}     
\usepackage{graphicx}    
\usepackage{xcolor}      
\usepackage{tabulary}    
\usepackage{tikz}        
\usetikzlibrary{shapes,arrows,decorations.pathreplacing,calligraphy} 
\usepackage{threeparttable}
\usepackage{svg}
\usepackage{rotating}
\usepackage{natbib}

\begin{document}
\begin{titlepage}
\begin{spacing}{1}
\title{\LARGE \textsc{Social Learning with Intrinsic Preferences}\footnote{Support from
the Deutsche Forschungsgemeinschaft (DFG, German Research Foundation) under Germany’s Excellence
Strategy EXC 2117–422037984 is gratefully acknowledged. We would like to thank Brendan Barrett, Yves Breitmoser, Sebastian Fehrler, Moritz Janas, Katrin Schmelz, the research group of the Cluster for the Advanced Study of Collective Behaviour, the Thurgau Institute of Economics (TWI), and the Department of Economics at the University of Konstanz, as well as participants of various conferences, workshops and seminars for helpful comments and suggestions. We are indebted to Charles Efferson for many insightful comments on an earlier draft of this paper. All remaining errors are our own.}
}
\author{\large{Fabian Dvorak}\vspace{2pt}\\ \normalsize{University of Konstanz} \vspace{2pt}\\ \small{fabian.dvorak@uni.kn}
\and
\large{Urs Fischbacher}\vspace{2pt}\\ \normalsize{University of Konstanz}\vspace{2pt}\\ \small{urs.fischbacher@uni.kn}
   }

\date{\small \today}

\maketitle
\thispagestyle{empty}
\vspace{-1cm}

\begin{abstract}
\noindent
Despite strong evidence for peer effects, little is known about how individuals balance intrinsic preferences and social learning in different choice environments. Using a combination of experiments and discrete choice modeling, we show that intrinsic preferences and social learning jointly influence participants' decisions, but their relative importance varies across choice tasks and environments. Intrinsic preferences guide participants' decisions in a subjective choice task, while social learning determines participants' decisions in a task with an objectively correct solution. A choice environment in which people expect to be rewarded for their choices reinforces the influence of intrinsic preferences, whereas an environment in which people expect to be punished for their choices reinforces conformist social learning. We use simulations to discuss the implications of these findings for the polarization of behavior.
\end{abstract}
\end{spacing}
  \vspace{10pt} 
\end{titlepage}

\section{Introduction}

Many decisions, such as how to dress, which sources of information to trust, and what to buy, are influenced not only by intrinsic preferences about choice alternatives, but also by social learning. This dual nature of decision making can lead to a fundamental conflict in situations where individual preferences differ from the behavior suggested by peers. Despite overwhelming evidence for the influence of both individual preferences and social learning on human decisions, little is known about how individuals trade off individual preferences and social learning in different choice tasks and choice environments.

Separating the influence of intrinsic preferences and social learning on people's decisions is notoriously difficult \citep{Manski1993}. People may behave similarly because they have similar intrinsic preferences for choice alternatives or because they are conformist social learners. At the same time, someone with unusual preferences may be misidentified as anticonformist when making independent decisions. To further complicate the identification of preferences and social learning, people tend to surround themselves with people who have similar preferences, making the information we receive about the behavior of others endogenous. Thus, to cleanly separate the effects of intrinsic preferences and social influence on individuals' decisions, we would need to measure preferences in the absence of social information and compare them to decisions made under exogenous social influence.

In this paper, we use a combination of experimental data and discrete choice modeling to cleanly separate the effects of intrinsic preferences and social learning on participants' decisions in two different choice tasks and three different choice environments. Our main contribution is the finding that preferences and social learning jointly influence participants' decisions in all experimental conditions, suggesting that models of human behavior should account for both factors. However, the relative importance of the two factors differs substantially and depends on both the choice task and the choice environment. Despite this complexity, a tractable pattern emerges. We find that intrinsic preferences are generally more important in a subjective choice task in which participants choose between works of art, whereas social learning determines participants' decisions in a task in which participants are instructed to choose the correct answer to a question. A choice environment in which people expect to be rewarded reinforces the influence of intrinsic preferences, whereas an environment in which people expect to be punished reinforces conformist social learning.

Combining experimental data with discrete choice modeling also allows us to identify participants' individual social learning strategies. We find that individual social learning strategies vary substantially across individuals, experimental tasks, and treatments. Conformist social learning is the norm in both experimental tasks, but is more common in the experimental task where participants choose answers to questions with an objectively correct answer. In the experimental task where participants choose one of two pieces of art, social learning is much more diverse and anticonformity occurs. When some participants are rewarded for their choices, individual social learning strategies cover the full spectrum of social learning in the painting selection task. Consistent with previous findings \citep{Dvorak2020}, conformist social learning dominates in the environment where participants expect punishment and becomes less frequent when participants expect reward. In the painting selection task, the average social learning strategy shifts from conformity in the control treatment to nonconformity in the reward environment, with implications for simulated population dynamics.

In addition to examining how people trade off intrinsic preferences and social learning, our study makes several methodological contributions. We extend the experimental design of \cite{Dvorak2020}, which allows us to predict participants' intrinsic preferences by using past choices of the same participant, to study social learning in groups larger than three. Predicting participants' choices mitigates the problem that participants may want to vote consistently, which would inflate estimates of the role of preferences in decision making. In the key phase of the experiment, we inform participants of the choices of five other (uninfluenced) participants in the same decision. We then use statistical modeling to disentangle the effects of intrinsic preferences and social influence on participants' decisions. The basis of our modeling approach is a random utility model in which discrete choices depend on (1) intrinsic preferences and (2) the expected frequency of others' choices. The model predicts several stylized facts about social learning, but also generates new insights into when and how social learning works. Perhaps most importantly, the interplay of preferences and social learning suggests that social norms are not arbitrary, but have some intrinsic value beyond the benefits of coordination.

In addition to integrating intrinsic preferences and social learning, our modeling framework extends existing models of social learning in two important ways. First, it goes beyond the simple threshold model of collective behavior \citep{Granovetter1978} by allowing for a variety of individual social learning strategies, including conformity, nonconformity, independence, and anticonformity. Second, the model incorporates prior beliefs about others' choices into the decision-making process, which are updated after observing a random sample of behavior. In the model, individual utility is a function of the individual's posterior belief about the frequency of choices in the reference group and \textit{not} a function of the frequency of choices in the sample. This is not only conceptually plausible, as it prevents small samples from triggering unusually strong social learning, but also avoids numerical problems in estimating social learning strategies at the individual level. Furthermore, explicit modeling of prior beliefs can be useful to incorporate biases in the decision process \citep{Benabou2002,Koszegi2006,Zimmermann2020}.

\section{Methods}
\label{sec:methods}

\subsection{Experiment}
\label{sec:experiments}

\subsubsection*{Choice tasks}
In the online experiment, participants make binary choices in two different choice tasks. The first choice task involves selecting one of two artistic paintings by well-known artists. The second choice task is to select the correct answer from two possible answers to a knowledge question. A list of all paintings, questions and answers can be found in the Tables \ref{tab:paintings}, \ref{tab:questions1} and \ref{tab:questions2} in the appendix.  

Each subject makes 60 binary decisions in each choice task. The first 50 decisions in each task are made without information about other participants' choices. In the last 10 decisions, the participant is informed about the choices of five other participants in the same decision. We use each participant's first 50 decisions to predict her uninfluenced preferences in the last 10 decisions with social information. 

To incentivize participants' decisions in the painting selection task, participants received the motif selected in one of the binary choices printed on a postcard. The postcard was mailed to the participant in an envelope a few days after the experiment. To increase intrinsic motivation to provide correct answers in the question task, participants know that they will be informed of the correct answers at the end of the experiment.

\subsubsection*{Measurement of intrinsic preferences}
The left panel of Figure \ref{fig:design} illustrates how we predict participants' preferences in the last 10 decisions of each choice task. We create sets that consist of four choice alternatives $A$, $B$, $C$, $D$. Depending on the task, these are four different paintings or four potential answers to the same questions. Each participant makes one binary choice for each possible pairwise combination of the four alternatives.

After each choices, the participant indicates the strength of her preference using a slider ranging from minus one (strong preference for the alternative displayed on the left), over zero (indifference) to one (strong preference for the alternative displayed on the right). The slider appears at the bottom of the decision screen after each decision. An example of the decision screen with the slider can be found in the Appendix. 

To predict the individual preference $\Delta$ of a participant in the binary choice between alternatives $A$ and $B$, we use the comparisons of both alternatives to the remaining alternatives $C$ and $D$. To predict the binary choice $A$ vs. $B$, we use the four \textit{related} binary choices $A$ vs. $C$, $B$ vs. $C$, $A$ vs. $D$ and $B$ vs. $D$, which are depicted in the left panel of Figure \ref{fig:design}.   

The self-reported preference strength in the two related choices that compare the alternatives $A$ and $B$ to a common third alternative yield an estimate for the participant's intrinsic preferences in the decision $A$ vs. $B$. Since there are two comparisons of $A$ and $B$ to a common third alternative, we use the mean of the two estimates as a measure for the participant's intrinsic preferences in the decision $A$ vs. $B$. 

In the example in the left panel of Figure \ref{fig:design}, the participant chooses alternative $C$ over $A$ and reports a preference strength of $0.5$. In the binary choice $B$ vs. $C$, the same participant prefers $B$ and reports a preference strength of $-0.5$. The difference between the two reported values divided by two produces an estimate for the intrinsic preference of the participant in the choice $A$ vs. $B$. The division makes sure that the estimate is between minus one and one. The two other related binary comparisons $A$ vs. $D$ and $B$ vs. $D$ yield an estimate of $0.3$. In the example, we would use the expected value $\frac{0.5+0.3}{2}$ of the two estimates as a measure for the intrinsic preference of the participant in the decision $A$ vs. $B$.

The sign of $\Delta$ indicates the preferred alternative of the participant. In the example, the sign of the expected value is positive which indicates that alternative $B$ is intrinsically preferred. The absolute difference in intrinsic utility between the two alternatives is measured by $\Delta$, the absolute value of the expected value. The variable $\Delta$ ranges between zero and one. In the example, the estimated absolute difference in  intrinsic utility between alternatives $A$ and $B$ is 0.4 - which is less than half of the maximum preference strength.

The method to infer intrinsic preferences illustrated in Figure \ref{fig:design} rests on the assumption of transitivity across the four choice alternatives. The method has the advantage that the participants of the experiment do not have to make the same binary choice twice - once without and once with social learning. This reduces the concerns that participants want to make consistent choices, which would lead to an underestimation of the effect of social learning. The technique to measure intrinsic preferences presented in the left panel of Figure \ref{fig:design} produces estimates that are proportional to the true intrinsic preferences plus some independent, identically distributed random error as long there is no systematic violation of transitivity. 

\begin{figure}[htbp]
\caption{Experimental design}
\label{fig:design}
\begin{center}
\fbox{
\begin{tikzpicture}
\node[align=left] (header) at (0.75,3.7) {\textbf{measurement of  $\Delta$}};
\draw (-1.3,3.4) -- (2.9,3.4);
\node[align=center] at (0,2.8) {\textit{related choices}};

\node[align=center] at (-1,2) {A}; 
\node[align=center] at (1,2) {\textbf{C}}; 
\node[align=center] at (-1,1) {\textbf{B}}; 
\node[align=center] at (1,1) {C}; 

\node[align=center] at (-1,0) {A}; 
\node[align=center] at (1,0) {\textbf{D}}; 
\node[align=center] at (-1,-1) {B}; 
\node[align=center] at (1,-1) {\textbf{D}}; 
 
\draw [lightgray] (-0.5,2) -- (0.5,2);
\draw [black] (0.25,1.9) -- (0.25,2.1);
\node[align=center] at (0.25,1.7) {{\scriptsize 0.5}};

\draw [lightgray] (-0.5,1) -- (0.5,1);
\draw [black] (-0.25,0.9) -- (-0.25,1.1);
\node[align=center] at (-0.25,0.7) {{\scriptsize -0.5}};

\draw [lightgray] (-0.5,0) -- (0.5,0);
\draw [black] (0.4,-0.1) -- (0.4,0.1);
\node[align=center] at (0.4,-0.4) {{\scriptsize 0.8}};

\draw [lightgray] (-0.5,-1) -- (0.5,-1);
\draw [black] (0.1,-0.9) -- (0.1,-1.1);
\node[align=center] at (0.1,-1.4) {{\scriptsize 0.2}};

\node[align=center] at (3,2.8) {$A$ vs. $B$};
\node[align=center] at (3,1.5) {$\frac{0.5 - (-0.5)}{2} = 0.5$}; 
\node[align=center] at (3.2,-0.5) {$\frac{0.8 - 0.2}{2} = 0.3$}; 

\node[align=center] (zerofour) at (4.05,0.5) {\textcolor{red}{$0.4$}};

\draw[->] (4.05,1.2) -- (4.05,0.8);
\draw[->] (4.05,-0.2) -- (4.05,0.2);

\draw [decorate,decoration = {brace}] (1.4,2.2) --  (1.4,0.8);
\draw [decorate,decoration = {brace}] (1.4,0.2) --  (1.4,-1.2);
\end{tikzpicture}
}
\hspace{-0.25cm}
\fbox{
\begin{tikzpicture}
\node[align=left] (header) at (1.1,3.7) {\textbf{sequence of decisions}};
\draw (-1.1,3.4) -- (3.4,3.4);
\node[align=center] (A) at (-1,-1.305) {\textcolor{white}{A}}; 
\node[align=center] at (0,2.8) {\textit{comparisons}};
\node[align=center] (A) at (-1,2) {A}; 
\node[align=center] (B) at (1,2) {B}; 
\node[align=center] (C) at (-1,0) {C}; 
\node[align=center] (D) at (1,0) {D}; 

\draw (A) -- (B) node[midway,above] {{\tiny C1}};
\draw (A) -- (C) node[midway,right] {{\tiny C2}};
\draw (A) -- (D) node[midway,right=8pt, above=10pt] {{\tiny C3}};
\draw (B) -- (C) node[midway,left=8pt, below = 10pt] {{\tiny C4}};
\draw (B) -- (D) node[midway,left] {{\tiny C5}};
\draw (C) -- (D) node[midway,below] {{\tiny C6}};

\node[align=center] at (1.8,2.8) {\textit{sbj}};
\node[align=center] at (1.8,2) {{\small 1}};
\node[align=center] at (1.8,1.6) {{\small 2}};
\node[align=center] at (1.8,1.2) {{\small 3}};
\node[align=center] at (1.8,0.8) {{\small 4}};
\node[align=center] at (1.8,0.4) {{\small 5}};
\node[align=center] at (1.8,0) {{\small 6}};

\node[align=center] at (3.6,2.8) {\textit{phase 1}};
\node[align=center] at (3.6,2.0) {{\footnotesize C1,C2,C3,C4,\textcolor{red}{C5}}};
\node[align=center] at (3.6,1.6) {{\footnotesize C1,C2,C3,C4,C6}};
\node[align=center] at (3.6,1.2) {{\footnotesize C1,C2,C3,\textcolor{red}{C5},C6}};
\node[align=center] at (3.6,0.8) {{\footnotesize C1,C2,C4,\textcolor{red}{C5},C6}};
\node[align=center] at (3.6,0.4) {{\footnotesize C1,C3,C4,\textcolor{red}{C5},C6}};
\node[align=center] at (3.6,0.0) {{\footnotesize C2,C3,C4,\textcolor{red}{C5},C6}};

\node[align=center] at (5.8,2.8) {\textit{phase 2}};
\node[align=center] at (5.8,2.0) {{\footnotesize C6}};
\node[align=center] at (5.8,1.6) {{\footnotesize \textcolor{red}{C5}}};
\node[align=center] at (5.8,1.2) {{\footnotesize C4}};
\node[align=center] at (5.8,0.8) {{\footnotesize C3}};
\node[align=center] at (5.8,0.4) {{\footnotesize C2}};
\node[align=center] at (5.8,0.0) {{\footnotesize C1}};

\end{tikzpicture}
}\\\vspace{-0.2cm}
{\setstretch{0.9}
\begin{flushleft}
{\footnotesize \textit{Notes: Left panel illustrates the method to measure the intrinsic preference a participant in the binary decision $A$ vs. $B$. The estimate $\Delta$ of the intrinsic preference is the expected value of two normalized estimates that are derived by comparing the two alternatives $A$ and $B$ to a common third alternative - either $C$ or $D$. The division by two normalizes the two estimates such that $\Delta$ ranges from -1 to 1. Right panel illustrates how the six pairwise  comparisons of four alternatives are distributed over two phases of the experiment. In phase 2, each participant is informed about the uninfluenced choices of the five other participants in phase 1.}}
\end{flushleft}
}
\end{center}
\end{figure}

\subsubsection*{Social information}
Before a participant makes the the decision with social learning between $A$ and $B$, the participant receives information about the decisions of five other participants in the same choice. Depending the experimental task, participants see the selected paintings or the selected answers of the five other participants on the decision screen. An example of the decision screen can be found in the Appendix.

At the time of their decisions, the five other participants themselves have no information about the decisions of other participants. The right panel of Figure \ref{fig:design} illustrates how the experimental design   efficiently makes sure that this condition is fulfilled for a group of six subjects. The pairwise comparisons of four alternatives $A$, $B$, $C$ and $D$ generate six binary decisions. In a first phase of the experiment, each participant of a group of six completes five of the six binary decisions without social information. In phase two of the experiment, the remaining sixth decision is completed by each participant. Since this sixth decisions differs for each participant of the group, it is possible to inform the participant about the choices of the other five participants from phase one. 

In the example in the right panel of Figure \ref{fig:design}, the decisions highlighted in red mark the decision C5 which is the comparison of alternatives $B$ and $D$. All subjects except subject 2 complete this decision in phase 1. Subject 2 can therefore be informed about the decisions of the five other subjects before she makes her decision in phase 2. The right panel of Figure \ref{fig:design} illustrates that this works for all six subjects. 
 
\subsubsection*{Environments with reward or punishment}
As in \cite{Dvorak2020}, we conduct two experimental treatments, one with reward and one with punishment, and a control treatment. In the treatments with reward or punishment, participants are informed that their decisions with social information, together with the choices of the five other participants, will be shown to a participant who is not part of the group of six. This participant takes the role of an evaluator. The six choices are displayed on the screen of the evaluator in a randomized order. The task of the evaluator, who does not know which of the displayed choices is the social choice, is to select one of the six participants by clicking on one of the six choices.  At the end of the experiment, one decision of the evaluator is randomly chosen for each experimental task. In the reward treatment, the payoff of the participant who was selected by the evaluator decision is increased by 10 Euro. In the punishment treatment, the payoff of the selected participant is reduced by 10 Euro. An example of the evaluation screen can be found in the Appendix \ref{app:study materials}. 

\cite{Dvorak2020} analyze the incentives in environments with reward or punishment based on a simple model of social evaluation in which the evaluator selects a participant based on her own intrinsic preference. This means the evaluator selects a participant who has chosen her preferred alternative for reward and a participant who has \textit{not} chosen her preferred alternative for punishment (if possible). They show that, with information about others behavior, the incentive to choose alternative $j$ depends the number $n_{ij}$ of other participants that have chosen alternative $j$ in phase 1 of the experiment. In the reward treatment, the incentive to choose alternative $j$ generally \textit{decreases} in $n_{ij}$. This incentivizes anti-conformity. In the punishment treatment, the incentive to choose alternative $j$ \textit{increases} in $n_{ij}$. This incentivizes conformity. The intuition behind this result is that, since the social evaluation is only based on partici
pants' choices, the probability of being rewarded or punished is evenly distributed among those who display the same behavior. Based on the theoretical analysis of \cite{Dvorak2020}, we expect that the choice environments with reward or punishment influence the model parameter $f$ that captures the reaction of a participant to $n_{ij}$ - the number of other individuals that have chosen alternative $j$. Compared to the control treatment, the social learning parameter $f$ should be larger under punishment and smaller under reward.

\subsubsection*{Implementation}

We conducted 10 experimental sessions online between December 11-16, 2020 with 174 students of various fields of the University of Konstanz in Germany. The mean age of the participants was 21.6 years. We recruited the participants of the experiment with the recruitment software \textit{hroot} \citep{Bock2014}. The experiments were conducted with \textit{z-Tree} \citep{Fischbacher2007} and \textit{z-Tree unleashed} \citep{Duch2020}. The duration of the experimental sessions was between 1.5 and 2 hours. The data set contains 3475 binary choices under social learning. Participants received a fixed amount of experimental currency for each decision in the experiment. On average, a participant earned 23 Euro for participating in the experiment. 

\subsection{Modeling framework}
Let $Y_i$ denote the discrete choice of individual $i$ among several alternatives with index $j \in \{1,\dots,J\}$. Let $U_{ij}$ represent the utility of alternative $j$ to individual $i$. We assume that the utilities $U_{ij}$ is the sum of two systematic components and one random component.

\begin{align}
\label{eq:utility}
U_{ij} =  \lambda_i v_{ij} + f_i \cdot log(s_{ij}) + \epsilon_{ij}
\end{align}

The two systematic components are the elements $\lambda_i v_{ij}$, and $f_i \cdot log(s_{ij})$ of the utility function. The component $\lambda_i v_{ij}$ reflects the \textit{intrinsic utility} of alternative $j$ for the individual. As in the conditional logit model \citep{McFadden1974}, it is assumed that intrinsic utility arises from several observable characteristics of the alternative multiplied by the individual marginal utilities of these characteristics. To simplify the notation, we consider a single characteristic $v_{ij}$ that yields marginal utility $\lambda_i$.

The second systematic component $f_i \cdot log(s_{ij})$ reflects the \textit{social utility} of the alternative. Social utility is a function of the natural logarithm of the expected share $s_{ij}$ of alternative $j$ in the reference group. The individual-specific parameter $f_i$ reflects individual $i$'s social-learning strategy, which is our main parameter of interest. The social-learning strategy of individual $i$ is conformist for $f_i > 1$, linear for   $f_i = 1$, and non-conformist $0 < f_i < 1$, independent for $f_i = 0$, and anti-conformist for $f_i < 0$.\footnote{As in frequency-dependent social learning, we define conformity as disproportional copying of behavior in form of an s-shaped response of the individual to the observed frequency of choices in the population~\citep{Efferson2008}. If all individuals are conformists, the s-shaped response implies polarization of behavior over time social learning goes beyond - which can explain a wide variety of behavioral phenomena ranging from the emergence and persistence of social norms \citep{Akerlof1980,Jones1984,Bernheim1994,Efferson2020} over inter-dependent consumer demand \citep{Gaertner1974,Pollak1976,Alessie1991} and neighborhood effects \citep{Schelling1971}, to the effectiveness of nudges and peer effects \citep{Allcott2011,Allcott2014,Alpizar2008,Bobek2007,Coleman2007,Nolan2008,Schultz2007,Smith2015}. } 

The component $\epsilon_{ij}$ captures randomness in the preferences of individual $i$ for the $j$th alternative. We will assume that the errors $\epsilon_{ij}$ are iid and follow a standard Gumbel distribution, which results in choice probabilities in the logit form \citep{Luce1959,McFadden1974}. The probability that individual $i$ chooses alternative $j$ has the following closed-form solution:
\begin{align}
\label{eq:response probabilities}
Pr(Y_i = j) = \frac{e^{ \lambda_i v_{ij} } (s_{ij})^{f_i} }{\sum_k e^{ \lambda_i v_{ik} } (s_{ik})^{f_i} }
\end{align}
Equation \eqref{eq:response probabilities} shows that the systematic components of the utility function influence the choice probabilities based on different functional forms. For the characteristics $v_{ij}$, the functional form is exponential. The exponential form has the advantage that the characteristics $v_{ij}$ can be specified in absolute or relative terms. 

Using the logarithm of the expected shares in the utility function \eqref{eq:utility} yields the power form for the expected shares. A practical advantage of the power form is that is invariant to multiplying all shares by a constant. As a result, the model makes the same predictions if one uses expected counts instead of expected shares, which is a convenient feature.

\subsubsection*{Beliefs about shares}
We assume that the share of individuals in the reference group that chooses alternative $j$ is ex-ante unknown. Individual $i$ holds a prior belief about the share of alternative $j$ that follows a Dirichlet distribution with parameters $\alpha =\alpha_{i1}, \dots,\alpha_{iJ}$. The $\alpha$-parameters can be interpreted as obseved counts and define the shape and the expected value of the prior belief  $s_{ij} = \alpha_{ij}/\sum \alpha_{ik}$ . If $\alpha_{ij} = 1 ~\forall j \in \{1,\dots,J\}$, the prior belief is uniform, i.e. all shares in the unit interval are believed to be equally plausible. In this case, the expected share is $1/J$. The larger $\alpha_{ij}$ is relative to the other $\alpha$-parameters, the more likely are large shares of alternative $j$. 

After observing a random sample of $N$ individuals, of which $n_{ij} \in \{0,\dots,N\}$ prefer alternative $j$, individual $i$ updates the prior belief about the share of alternative $j$ in the reference group. Since the Dirichlet distribution is the conjugate prior of the multinomial distribution, the resulting posterior also follows a Dirichlet distribution with updated parameters $\alpha_{ij} + n_{ij}$. After the belief update, the expected shares of the alternatives are 

\begin{align}
\label{eq:update}
s_{ij} = \frac{\alpha_{ij} + n_{ij}}{\sum^{J}_{k}\left( \alpha_{ik} + n_{ik} \right) }.
\end{align}

An important remark is that the utility defined in Equation \eqref{eq:utility} differs from the expected utility of the alternative given the posterior belief of the individual. Hence, Equation \eqref{eq:utility} suggests that the individual does \textit{not} form an expectation about the social utility of the alternatives. Instead, it models the following sequential cognitive process: the individual first updates the her belief when exposed to social information, calculates the expected share of each alternative in the reference group, and then decides based on the utilities of the alternatives that result from the expected shares.\footnote{The advantage compared to expected utility maximization is that $s_{ij}$ can be interpreted as the share of alternative $j$ in Equation \eqref{eq:utility}. Expected utility maximization requires that the individual maximizes $E(log(X))$ where $X$ is a Dirichlet distributed random variable with parameters $\alpha_{ij} + n_{ij}$ and:
\begin{align*}
E(log(X)) = \psi(\alpha_{ij} + n_{ij}) - \psi \left(\sum_k \alpha_{ik} + n_{ik} \right)
\end{align*}
where $\psi(\cdot)$ is the digamma function which has the logistic approximation $\psi(x) \approx log \left(x - \frac{1}{2} \right)$ \citep{Johnson1994}. Using this approximation: \begin{align*}
E(log(X)) \approx log \left( \frac{\alpha_{ij} + n_{ij} - \frac{1}{2}}{\sum_k ( \alpha_{ik} + n_{ik} ) - \frac{1}{2}} \right) \approx s_{ij}
\end{align*} if the quantity $\alpha_{ij} + n_{ij}$ is large for all alternatives.}
 
\subsubsection*{Biased beliefs}
\label{subsubsec:biased beliefs}
An advantage of explicitly modeling individuals' beliefs is that the model can be extended to account for biases in the belief-formation process. From Equation \eqref{eq:utility} it is clear that an individual might be in a situation in which her intrinsic preferences and social learning suggest different alternatives. For example, consider a conformist individual who strongly prefers one alternative but, at the same time, believes that the other alternative is more frequent in the reference group. Depending on the strength of her intrinsic preferences and the strength of social learning, the individual will either stick to the intrinsically preferred alternative at the cost of less social utility or switch to the more popular alternative at the cost of less intrinsic utility. One way to resolve such conflicts is by manipulating beliefs in a way that the social utility penalty for the intrinsically preferred alternative is less severe. Such acts of self-deception are known as false-consensus effect \citep{Mullen1985} or motivated beliefs \citep{Benabou2006}.

One possibility to incorporate motivated beliefs is to define the parameters of the prior belief distribution as:
\begin{align}
\label{eq:biased_prior}
\alpha_{ij} = \phi_i \cdot J \cdot \frac{e^{ \delta_i v_{ij} }}{\sum_k e^{ \delta_i v_{ik}}}
\end{align}
where $\phi_i > 0$ is the strength of prior belief and $\delta_i$ is a parameter for the preference bias. If the parameters $\delta_i$ and $\lambda_i$ have the same sign, the prior is biased such that large shares of the intrinsically preferred alternative are more likely. If the signs of the two parameters $\delta_i$ and $\lambda_i$ differ, the prior is biased such that small shares of the intrinsically preferred alternative are more likely. If $\delta_i$ is zero, the individual's prior belief is unbiased and the expected shares are $1/J$ for every alternative.

Other variants of preference-biased prior beliefs are also plausible.\footnote{An alternative way to introduce preference-biased beliefs would be to assume that the updating process of the individual is biased by her intrinsic preferences \citep{Zimmermann2020}. This can be achieved by introducing updating weights proportional to the intrinsic utility of each alternative.} We use the form outlined in Equation \eqref{eq:biased_prior} because of the convenient property that the prior expected share is:
\begin{align*}
s_{ij} = \frac{\alpha_{ij}}{\sum_k \alpha_{ik}} = \frac{e^{ \delta_i v_{ij} }}{\sum_k e^{ \delta_i v_{ik}}}\\
\end{align*}
If $\delta_i \approx \lambda_i$, this means that the expected prior belief corresponds to the own behavior in the absence of social learning.  

\subsubsection*{Comparative statics}
\begin{figure}
\caption{Predicted response to social learning}
\label{fig:response}
\vspace{0.2cm}
\includegraphics[scale=0.59]{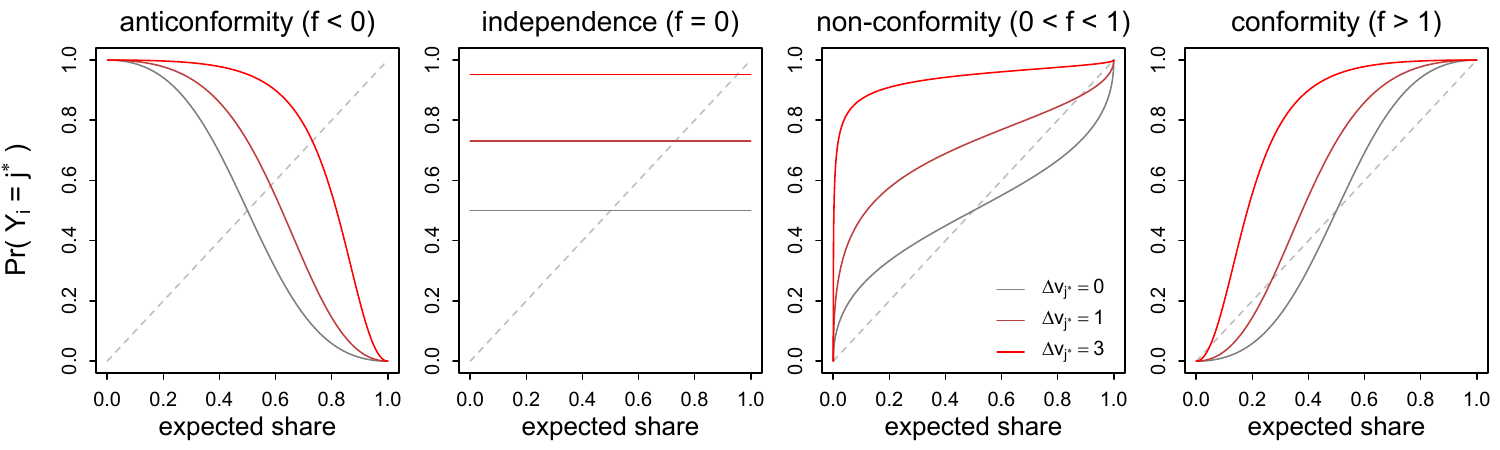}
\begin{tabulary}{15cm}{J}
		\setstretch{0.7}
		{\scriptsize \textit{Notes:} The four graphs show the probability of choosing the intrinsically preferred alternative $j^{*}$ in a binary choice as a function of the expected frequency of the preferred alternative $j^{*}$ in the reference group ($s_{j^{*}}$). From left to right, the graphs show this probability for an anticonformist, independent, non-conformist and conformist individual, characterized by the social learning parameter ($f$). The solid lines assume that $v_{j}$ is a dummy for the preferred alternative and vary the $\lambda$ parameter. 
\par} 	
	\end{tabulary}
\end{figure}

The modeling framework conveys several stylized facts about social learning. The four graphs of Figure \ref{fig:response} illustrate the effects of the parameters $\lambda$ and $f$ on the probability to choose the preferred alternative. We use the index $j^{*}$ to indicate the preferred alternative. The probability to choose alternative $j^{*}$ is a function of the expected share of alternative $s_{j^{*}}$ in the reference group depicted on the x-axis. In Figure \ref{fig:response}, this function is plotted for a binary choice. Note that we omit the individual index $i$ for better readability. 

The four different panels illustrate the shape of the probability function for different values of the social learning parameter $f$. Conditional on the value of the social learning parameter, the individual is classified as anticonformist, independent, non-conformist or conformist. While anticonformity (left panel) implies that the probability to choose the preferred alternative decreases in the expected frequency of the preferred alternative in the reference group, non-conformity and conformity (central-right and right panels) both imply a positive relation between the probability to choose alternative $j^{*}$ and the expected share of $j^{*}$ in the reference group. Independence (central-left panel) implies that the probability to choose alternative $j^{*}$ is not affected by the expected frequency of the alternative in the reference group. 

The solid red lines in Figure \ref{fig:response} show how the probability to choose alternative $j^{*}$ is influenced by the model parameter $\lambda$. The more red the color of a line is, the stronger is the utility difference between the alternatives. Figure \ref{fig:response} illustrates that intrinsic preferences bend the probability curves upwards which makes it more likely that the individual chooses the intrinsically preferred alternative.

Figure \ref{fig:response} illustrates three stylized facts about social learning:
\begin{enumerate}
\item \textit{Intrinsic preferences immunize against social information.} Figure \ref{fig:response} shows how intrinsic preferences mediate the response to social information. The key aspect to observe is that intrinsic preferences for alternative $j^{*}$ (solid red lines) create a larger region of possible beliefs for which the predicted response curve is relatively high and flat, i.e. insensitive to changes in the belief. This implies that (1) the individual almost certainly chooses the intrinsically preferred alternative $j^{*}$ and (2) needs a relatively large and homogeneous sample to overcome this tendency. 
\item \textit{Strong prior beliefs immunize against social information.}
Figure \ref{fig:response} also illustrates the role of prior beliefs for individuals choices. Before observing a sample of others' behavior, the choice probability is determined the expected share of the preferred alternative in the reference group. When observing the sample, the individual updates the prior belief. If the observed frequencies differ from the individual's prior belief, the expected share will also differ and the choice probability will change accordingly. The strength of the individual's prior belief defines the distance between the prior and the posterior expected share on the x-axis of Figure \ref{fig:response}. If choices are not independent ($f \neq 0$), a strong prior implies a small change in the choice probability before and after observing the sample. Conversely, this means that, given some strength of the prior belief ($\phi > 0$), the effect of social information increases in the size of the sample. 

\item \textit{Social norms are not arbitrary.} Each response curve depicted in Figure \ref{fig:response} implies different choice dynamics in a group of repeatedly interacting individuals. The intersections of the solid lines with the dashed lines fix points of the choice dynamics, assuming that all individuals are the same and individuals' beliefs will eventually converge towards the true choice frequencies. For $f < 1$ the red lines suggest converges to a state in which the majority of individuals chooses the intrinsically preferred alternative. In this state of convergence, the frequency of the majority behavior is indicative of the strength of the intrinsic preference for the majority choice. For conformity ($f >1$) the strength of the intrinsic preference increases the basin of attraction of the majority choice which means that the group converges to the intrinsically preferred choice with higher probability. Independent of the type of social learning, the behavior of the group contains information about intrinsic preferences of the group after convergence. We run simulations to assess whether the behavior of the majority contains also information about the average preferences in a population with heterogeneous preferences. Figures \ref{fig:simulation} and \ref{fig:simulation_minority} in the Appendix show that this is the case for a broad range of model parameters, and also, if a minority with strong preferences against the majority behavior exists. 
\end{enumerate}

The model nests other models of collective behavior. Frequency-dependent social learning model \citep{Boyd1985} is obtained for $\lambda_i := 0$ and $\phi_i := 0$. The influential threshold model of collective behavior \citep{Granovetter1978} is obtained by assuming $f_i \rightarrow \inf$ and $\phi = 0$.

We fit Bayesian multilevel models to the data of the online experiment using \textit{R} \citep{R2021}, \textit{Stan} \citep{Stan2021} and the packages \textit{RStan} \citep{RStan2020} and \textit{rethinking} \citep{McElreath2020}. Details about the Bayesian parameter estimation can be found in the Appendix.

\section{Results} 
\label{sec:results}

\subsection{Population parameters}
Table \ref{tab:post dist} shows estimated population means of the model parameters along with the 5th and 95th percentiles of their posterior distributions. For the social learning parameter $f$, the table shows the estimates for the control treatment, without reward or punishment. The posterior means of all population parameters are larger than zero which suggests that participants' decisions are generally influenced by intrinsic preferences, social learning and preference-biased prior beliefs. Table \ref{tab:post dist} also shows that the parameter estimates vary substantially between the two experimental tasks.

\begin{table}[htbp]
\caption{Population means of model parameters}
\label{tab:post dist}
\centering
\begin{tabular}{lcccccccccccc}
  \hline \hline \\[-2ex]
  & \multicolumn{2}{c}{$\lambda$} & \multicolumn{2}{c}{$f$} & \multicolumn{2}{c}{$\phi$} & \multicolumn{2}{c}{$\delta$}  \\
  \cline{2-3} \cline{4-5} \cline{6-7} \cline{8-9}
 task & quest & paint & quest & paint & quest & paint  & quest & paint\\ 
  \hline
mean & 2.89 & 5.09 & 2.61 & 1.22 & 3.29 & 1.98 & 2.89 & 2.36 \\ 
5th percentile & 0.99 & 3.25 & 1.81 & 0.57 & 1.66 & 0.48 & 1.47 & 0.32 \\ 
95th percentile & 4.72 & 6.75 & 3.55 & 2.13 & 5.72 & 5.93 & 4.36 & 4.18 \\ 
  \hline \hline \\[-2ex]
\end{tabular}
{\setstretch{0.9} 
\begin{flushleft}
{\footnotesize \textit{Notes: Table shows means and percentiles of the posterior distribution of the model parameters. The values for the social learning parameter $f$ are reported for the control treatment.}}
\end{flushleft}
}
\end{table}

The posterior means of the parameter $\lambda$ indicate that the measured intrinsic preferences influence the participants' decisions. All estimates are clearly positive, indicating a tendency to choose the alternative with greater intrinsic utility.  The posterior means of the social learning parameter $f$ indicate conformity in both experimental tasks for the control treatment. The estimates indicate more conformity in the questioning task. The estimates of the belief parameters $\phi$ and $\delta$ suggest that participants hold preference-biased prior beliefs. The estimates of $\delta$ are positive. On average, participants believe that it is more likely that their intrinsically preferred alternative will be chosen by more individuals in the reference group.

\subsection{Variation of social learning and beliefs}
\begin{figure}[htbp]
\caption{Individual social-learning strategies and beliefs}
\label{fig:individual}
\begin{center}
\includegraphics[scale=0.7]{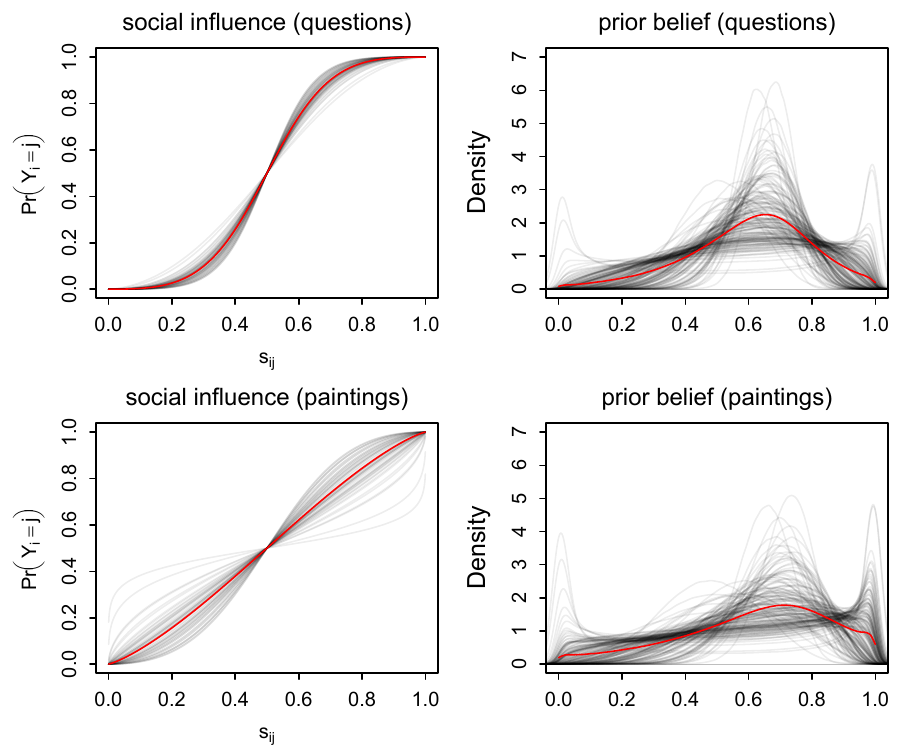}
\end{center}
{\setstretch{0.9} 
\begin{flushleft}
{\footnotesize \textit{Notes: Figure depicts individual estimates for social learning (left graphs) and prior beliefs (right graphs). Red lines indicate the average of the individual estimates.}}
\end{flushleft}
}
\end{figure}

Figure \ref{fig:individual} depicts individual heterogeneity in social learning (left graphs) and prior beliefs (right graphs) for the two experimental tasks. Each line represents an individual estimate. The red curves indicate the mean of the individual curves.   

The upper left graphs of Figure \ref{fig:individual} suggests that participants' conformity prevails in the questions task with little individual heterogeneity. The inter-individual differences are more pronounced in the painting selection task. In this task, the majority of individuals is conformist and displays an S-shaped reaction to the expected frequency of an alternative. A minority is non-conformist with an inverse S-shaped reaction. 

The right graphs depict individuals' prior beliefs about the frequency of their intrinsically preferred alternative in the reference group. The graphs suggest that majority of individuals discount the information presented to them about the frequency of choices. Some individuals hold a U-shaped prior which suggests no discounting of social information. The depicted prior beliefs also show that most individuals hold the preference-biased prior belief that larger frequencies are more plausible for their intrinsically preferred alternative. 

\subsection{Social learning under reward and punishment}
As in \cite{Dvorak2020}, we find that the evaluators allocate punishment and reward based on their own intrinsic preferences. Based on their theory and experimental findings, we expect that punishment induces conformity while reward induces anticonformity. This means that, compared to the control treatment, the social learning parameter $f$ should be larger under punishment leading to more conformist social learning and smaller under reward leading to less conformist social learning. 

\begin{figure}[htbp]
\caption{Individual social-learning strategies}
\label{fig:treatment}
\begin{center}
\includegraphics[scale=0.7]{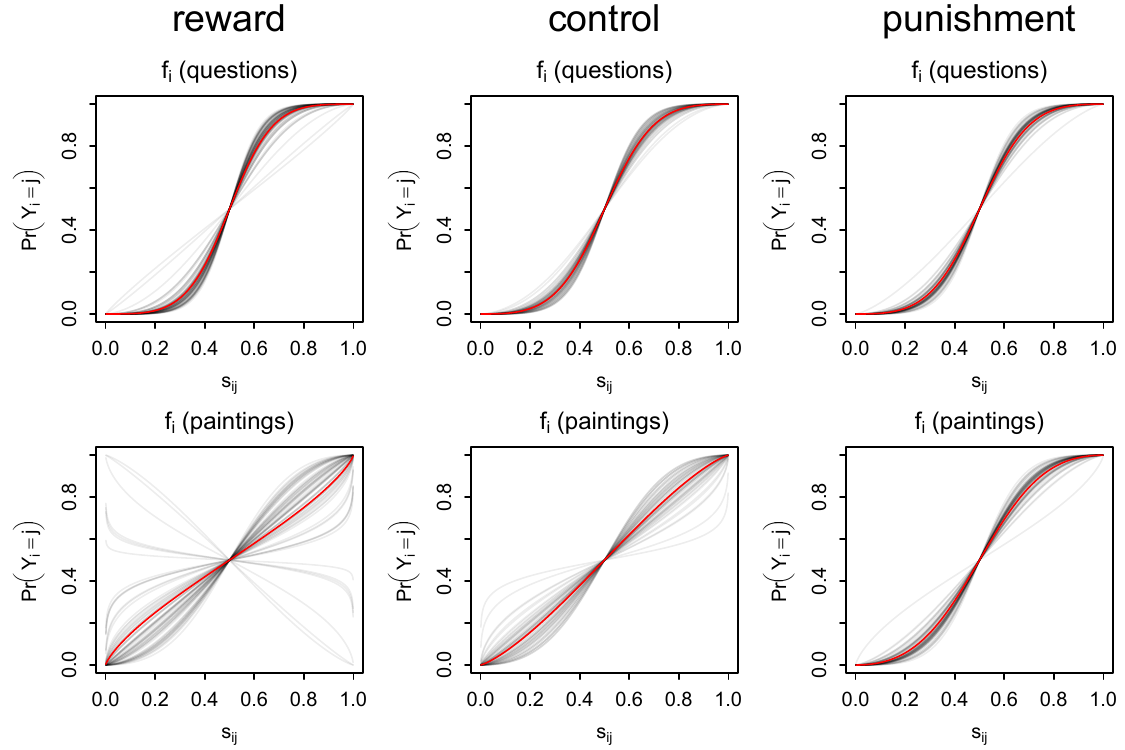}
\end{center}
{\setstretch{0.9} 
\begin{flushleft}
{\footnotesize \textit{Notes: Individual social-learning strategies for the two experimental tasks (rows) in different choice environments (columns). Red lines indicate the average social-learning strategy.}}
\end{flushleft}
}
\end{figure}

Figure \ref{fig:treatment} illustrates the effects of the reward and punishment treatment on average (red lines) and individual (grey lines) social learning strategies. For the painting task, represented by the three panels in the second row, the effects are in line with our prior expectations. Social learning is indeed less conformist (negative effect on $f$) in the reward environment and more conformist (positive effect on $f$) in the environment with punishment. In the reward treatment (left panel), the average behavior indicated by the red line is non-conformist, and a small fraction of individual estimates even indicates anticonformity $f<0$. The average social learning strategy almost linear in the control treatment (center panel), with a slight conformist twist. In the environment with punishment, average social learning is clearly conformist. This suggests that environments with reward can induce different choice dynamics compared to environments with punishment, in which standing out makes little sense. 

For the questions task, the treatments effects are not consistent with prior expectations. The different treatments have no effect on social-learning, which is strongly conformist in all three treatments.

\subsection{Model fit and model comparisons}
\begin{figure}[htbp]
\caption{Fit for average and individual behavior}
\label{fig:data}
\begin{center}
\includegraphics[scale=0.7]{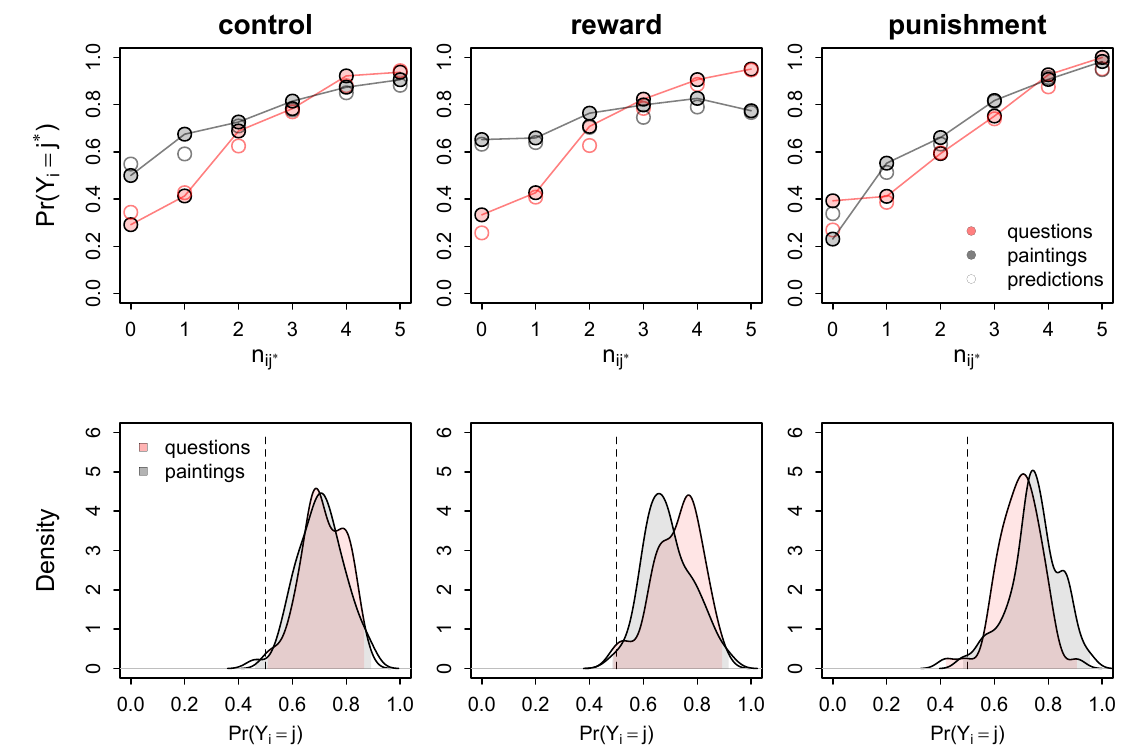}
\end{center}
{\setstretch{0.9} 
\begin{flushleft}
{\footnotesize \textit{Notes: Upper part shows model fit to average behavior. Circles represent model predictions. Dots represent average behavior observed conditional on the information about others behavior. Lower part shows the distribution of the mean posterior probability of all the choices from the same participant.}}
\end{flushleft}
}
\end{figure}

A natural question is how well the estimated model fits the data, and how the model fit compares to models with fewer parameters. Figure \ref{fig:data} illustrates the model fit for average and individual behavior. The three plots in the first row of Figure \ref{fig:data} plot model predictions (circles) against observed average behavior (dots) for the two experimental tasks in each of the three experimental treatments. The filled dots depict the average probability that a participant chooses her intrinsically preferred alternative $j^{*}$ conditional on $n_{ij^{*}}$ - the number of other participants that have chosen this alternative. The three plots illustrate why it is virtually impossible to infer the average social-learning strategy from average behavior. The model predictions are fairly close to the observed average behavior in the two experimental tasks of each treatment. The fit is generally better for larger values of $n_{ij^{*}}$. There seems to be no systematic error in the model predictions.

To illustrate the fit of the model to the individual behavior, we consider all choices from the same individual in the same task and calculate the mean posterior probability of these choices for different sets of model parameters sampled from the posterior distributions. The three graphs in the lower part of Figure \ref{fig:data} depict the distribution of the mean posterior probability of participants' choices conditional on the experimental task and treatment. In all three graphs, most of the density mass of is above one-half - which is the expected benchmark for random predictions indicated by the dashed line. Only few individuals fall below this benchmark. The mode of the mean posterior probability is around 70\%.

We compare the model to three nested models. The first model does not include a parameter for the preference-bias (PSI model). Instead, it uses $s_{ij} = (\phi_{it}+n_{ij})/(2\phi_{it} + 5)$. The second nested model is a pure preference model that only takes the intrinsic preferences over the choice alternatives into account (P model). The third model is a pure social learning model in which participants' choices depend on the expected frequencies of the choice alternatives (SI model). The three nested models are obtained from the preference-biased social learning model (PBSI model) outlined in Section \ref{sec:statistics} for the restrictions $\delta := 0$, $f := 0$ and $\delta := 0 \wedge \lambda := 0$ respectively.

Table \ref{tab:post means} summarizes the posterior means of the model parameters for all four models. The posterior means of the preference parameter $\lambda$ are larger in the PSI and P models and do not differ between the experimental tasks. The posterior means of the social learning parameters $f$ estimated for the control treatment are smaller in the PSI and SI models and suggest a non-conformist reaction to social information in the painting selection task. The estimates of the parameter $\phi$ - which indicates prior experience - are larger in these models. 

\begin{table}[htbp]
\caption{Posterior means}
\label{tab:post means}
\centering
\begin{tabular}{lcccccccccccc}
  \hline \hline \\[-2ex]
  & \multicolumn{2}{c}{$\lambda$} & \multicolumn{2}{c}{$f$} & \multicolumn{2}{c}{$\phi$} & \multicolumn{2}{c}{$\delta$} & WAIC \\
  \cline{2-3} \cline{4-5} \cline{6-7} \cline{8-9}
 task & quest & paint & quest & paint & quest & paint  & quest & paint \\ 
  \hline
PBSI & 2.89 & 5.09 & 2.61 & 1.22 & 3.29 & 1.98 & 2.89 & 2.36 & 3157  \\ 
PSI & 6.28 & 6.81 & 2.08 & 0.58 & 1.55 & 0.29 & - & - & 3180  \\
P & 6.36 & 6.68 & - & - & - & - & - & - & 3652 \\
SI & - & - & 1.94 & 0.69 & 1.70 & 0.39 & - & - & 4192 \\
  \hline \hline \\[-2ex]
\end{tabular}
{\setstretch{0.9} 
\begin{flushleft}
{\footnotesize \textit{Notes: Table compares posterior means of the model parameters of four different models. The last column depicts the value of the widely applicable information criterion \citep{Watanabe2010} for each model.}}
\end{flushleft}
}
\end{table}

The last column of Table \ref{tab:post means} indicates the values of the widely applicable information criterion \citep[WAIC,][]{Watanabe2010} for the four models.\footnote{
\begin{footnotesize}
The formula of the WAIC is:
\begin{align*}
WAIC(Y|\Psi) = -2\left( \sum_m log \frac{1}{S}\sum_s Pr( Y_m|\Psi_s ) - \sum_m var \left(log \left(Pr( Y_m | \Psi_s )\right) \right) \right)
\end{align*}
where $Y$ are the observed choices and $\Psi$ are the samples used to approximate the posterior \citep{McElreath2020}.
\end{footnotesize}} The WAIC approximates the out-of-sample prediction accuracy of a Bayesian model. It is asymptotically equivalent to leave-one-out cross-validation \citep{Vehtari2017}. Hence, models with low WAIC values are preferable. The WAIC values indicate that the model with intrinsic preferences, social learning and preference-biased prior beliefs generates more accurate out-of-sample predictions than the nested models.

We also fit models with fixed effects of the treatment condition for all model parameters. We find that these models do not yield better out of sample predictions according to the WAIC. This means that the treatment effect is best explained by the social learning parameter. Removing the task fixed effect of any model parameter also results in larger values of the WAIC.

\subsection{Simulation of choice dynamics}
The fitted model can be used to simulate  choice dynamics of the binary choices used in our experiment. We can go beyond the limitations of our experimental setup and explore choice dynamics in larger groups and over a longer time horizon. For the simulation of choice dynamics, we will have to assume that individual model parameters and intrinsic preferences do not change over time or when interacting in larger groups. It is important to stress that these assumptions cannot be checked with data of our experiment. For the results presented in the following, the reader should keep this limitation in mind.

\begin{figure}[htbp]
\caption{Simulation of choice dynamics}
\label{fig:dynamics}
\begin{center}
\includegraphics[scale=0.7]{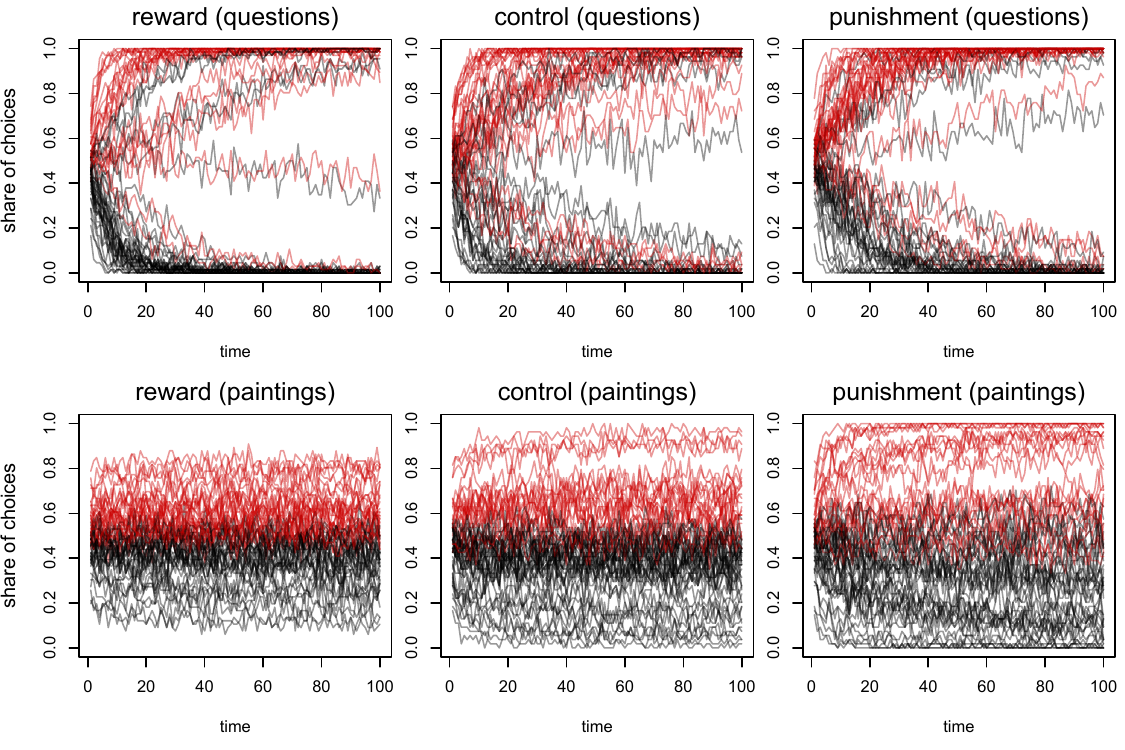}
\end{center}
{\setstretch{0.9} 
\begin{flushleft}
{\footnotesize \textit{Notes: Simulation of choice dynamics for the 120 unique binary choices using fitted individual choice parameters. After each period, each participant is informed about the decisions of five other participants in the same binary choice, randomly sampled from the same choice environment. Each line represents one binary choice and shows the evolution of the frequency of one of the two alternatives, the target alternative, over 100 periods. Red lines indicate an average preference for the target alternative.}}
\end{flushleft}
}
\end{figure}

Figure \ref{fig:dynamics} shows the simulated choice dynamics for the 120 unique binary choices of the online experiment over 100 periods. In each period, we simulate a choice for each participant based on the measured intrinsic preferences and the expected frequencies of others' decisions. After each period, each participant is informed about the decisions of five participants randomly samples from the same choice environment in the same choice. To simulate participants' choices and update their beliefs, we use the posterior means of individual parameter estimates. 

Each line in the six panels of Figure \ref{fig:dynamics} shows the choice dynamics of one binary choice. On the vertical axis, we plot the share of one of the two alternatives, the \textit{target alternative}. For the questions task, the target alternative is the correct answer to the question. For the painting selection task, we randomly select one of the two paintings that then represents the target alternative for every participant. The color of each line reflects the average preferences of the interacting participants in each panel. A red line indicates an average preferences for the target alternative. A black line indicates the opposite, i.e. average preferences for the other alternative.

The panels for the questions task in the first row of Figure \ref{fig:dynamics} show many instances of convergence of opinions after 100 periods. The panels also show considerable sorting according to the average preferences of the participants. Red lines tend to converge to a target share of one, and black lines tend to converge to target share of zero. This is particularly the case for choices that converge early. In the painting selection task depicted in the lower three panels, the fitted model parameters suggest few instances convergence after 100 periods and the effect of the choice environment on the diversity of opinions is clearly visible. Sorting takes place nevertheless as red lines tend to lie above one-half, and black lines below one-half, which indicates that the majority choice corresponds to the average preferences in the population. 

In summary, the results of the simulation suggest that the descriptive norm after several periods of social interaction can be used to infer the average preferences of the participants. This is true for both choice tasks and all three choice environments.

\section{Discussion}
\label{sec:discussion}
Understanding individual social learning from observational data is notoriously difficult. Manski~(\citeyear{Manski1993}) highlights the fundamental problem of separating social learning from correlated effects, i.e., the fact that people behave similarly because they have similar intrinsic preferences over choice alternatives. Thus, to isolate intrinsic preferences from social learning, one must control for individuals' intrinsic preferences. This raises the problem that in observational settings, individuals' intrinsic preferences are often unknown and can only be inferred from their choices - leading to a fundamental identification problem.

In this paper, we use a combination of experimental data and discrete choice modeling to understand the complexity and determinants of the interplay between preferences and social learning. We show that preferences and social learning jointly influence participants' decisions across experimental conditions, suggesting that models of human behavior must account for both factors. Participants follow a diverse set of social-learning strategies that are shaped by features of the choice task as well as features of the choice environment. The individual variation we find in social learning strategies is important because it goes beyond different facets of conformity. At the same time, contextual variation in social learning strategies follows a clear pattern. Social learning is predominantly conformist in tasks with an objectively correct solution. This changes little when incentives for anticonformity are introduced. In contrast, individuals display a variety of social learning strategies in tasks that are subjective and do not have an objectively correct solution. In such tasks, the incentives provided by the environment play a crucial role in the nature of social learning strategies, shifting toward conformism in punishment environments and toward nonconformism in reward environments.

Methodologically, we derive these results based on a modeling framework that is inspired by different disciplines: we combine random utility theory developed in the economics literature with the idea of frequency-dependent social learning which is used in psychology, biology and behavior ecology to study collective behavior in humans and other animals \citep{Boyd1985,McElreath2005,McElreath2008,Efferson2008,Aplin2017,Barrett2017,Toyokawa2019,Deffner2020,Toyokawa2021}. The advantage of the combined model is that it can disentangle the role of individual preferences and beliefs and social learning.

Our modeling approach differs in several important ways from the influential threshold model of collective behavior \cite{Granovetter1978}, which can be considered the standard approach to integrating individual variation in preferences and conformity. Interpreted as a model of intrinsic preferences plus conformity, the threshold model assumes that each individual can be characterized by an individual choice frequency threshold for following the majority. Individuals with strong minority preferences are characterized by large thresholds because they face higher intrinsic costs of conformity. The equilibria of collective behavior are defined by the cumulative distribution function of thresholds in the population. Our modeling framework allows for a more detailed analysis of individual social learning, featuring linear and nonlinear responses to social influence, and also accounts for variation in individual beliefs about the behavior of others. 

Beliefs are important when people have prior experience with others' behavior or extrapolate from similar experiences, but also if the individual cares about the entire reference group and only observes random sample. A similar argument applies if others' decisions are subject to errors and observations can thus be surprising because they are not in line with the individual's prior belief. The seminal studies of Asch (\citeyear{Asch1952,Asch1956}) on conformity provide a good example. One explanation why most participants did \textit{not} conform to the wrong majority opinion is that it was in strong contrast to their prior expectation about the majority opinion. Asch's finding that the number of the majority substantially increases conformity to the majority can also be interpreted along those lines. Prior experience plays an important role in field experiments, and particularly, in natural field experiments on peer influence.

The modeling framework used in this study can explain several stylized facts about social learning. The first concerns the limits of social learning. The model suggests that those who do not learn socially are either idealists with strong intrinsic preferences or strong believers with substantial prior knowledge about peer behavior. Idealists do not respond to social information because their intrinsic preference is too strong for social learning. Strong believers, on the other hand, are unresponsive to social information because their prior belief is too strong to be significantly affected by information about others' behavior. Both these mechanisms can also differences in individual behavior across situations.

The model can also explain the stylized fact that conformity increases with the size of the majority - a finding that pervades the experimental literature on conformity~\citep{Bond1996,Sasaki2019}. The reason is that, given the same prior belief, large samples have a greater effect on the individual's posterior belief.\footnote{Prior experience can also explain social inertia in behavior which is sometimes observed when social learning is supposed to induce social change. If the status-quo exist for a long time, individuals develop strong beliefs that, in combination with conformity, stabilize the status-quo \citep{Smerdon2020}. Another advantage of explicitly modeling individuals' beliefs is that models of social learning can be extended to incorporate biases in the formation of beliefs \citep{Benabou2002,Koszegi2006,Zimmermann2020}. Such biases seem plausible since individuals frequently face a trade-off between intrinsic preferences and social learning and one way to resolve the conflict is to engage in self-deception. For example, a conformists could overestimate the frequency of her preferred alternative to resolve the conflict that arises when she intrinsically prefers the minority choice. 
} This also has an econometric advantage. When fitting frequency-dependent social learning models to the decisions of individuals influenced by small samples, the models predict some decisions with certainty because observing small homogeneous samples generates exceptionally strong social learning. A single decision in the data that does not match the model's prediction means that the model has a likelihood of zero, making it impossible to estimate parameters. Prior beliefs effectively reduce the impact of small samples and circumvent this problem.

Finally, and perhaps most importantly, the modeling framework developed in this study can explain why it can be reasonable to infer intrinsic preferences from average behavior even when individuals are social learners who have interacted in the past. The reason is that the basin of attraction of the behavior preferred by the majority is larger. If all individuals are conformists and have the same intrinsic preferences, the group's choices are more likely to converge on the intrinsically preferred behavior. If the individuals are nonconformists, the choices will not converge, but the intrinsically preferred behavior will still be performed more often. Thus, the model supports the view that social norms are not arbitrary, but have some intrinsic value beyond the mere benefits of coordination.

\bibliographystyle{ecta}
\bibliography{koNtiki}

\newpage
\appendix

\section*{Appendix}

\subsubsection*{Bayesian multilevel modeling}
\label{sec:statistics}
We fit Bayesian multilevel models to the data of the online experiment using \textit{R} \citep{R2021}, \textit{Stan} \citep{Stan2021} and the packages \textit{RStan} \citep{RStan2020} and \textit{rethinking} \citep{McElreath2020}. The advantage of the multilevel specification is that the amount of information pooling between individual parameter estimates is defined by the data. This effectively prevents over-fitting of individual data, which would lead to overestimation of the variation in social-learning strategies and other individual differences. 

The multilevel models include individual random effects (index $i$) and fixed effects for the experimental task (index $t$) for all model parameters. Importantly, the random effects are drawn from a common distribution characterized by a population mean and some finite population variance. We introduce additional fixed effects for the effect of the treatment conditions (index $c$) on the social learning parameter $f$ to estimate the effect of the experimental treatment on this model parameter. Hence, we assume that the effects of the choice task and the choice environment have a similar effect on all individuals and do not affect the individual variation of social learning.

We use standard priors that result in expected effects of zero which corresponds to random choice and Hamiltonian Monte-Carlo \citep{Neal2011,Betancourt2013} to approximate the posterior distributions of the parameters.\footnote{\begin{small} We use the standard normal distribution for the prior of the fixed effects of all model parameters except $\phi$ which cannot be negative. For the parameter $\phi$, we use $exp(N(0,1))$. As prior for the individual varying effects, we use the multivariate normal distribution $MVN(0,\Sigma)$ with:
\begin{align*}
\Sigma =  \begin{pmatrix}
\sigma_{\lambda} & \dots & 0 \\
\vdots & \ddots & \vdots \\
0 & \dots & \sigma_{\delta} 
\end{pmatrix} \Omega \begin{pmatrix}
\sigma_{\lambda} & \dots & 0 \\
\vdots & \ddots & \vdots \\
0 & \dots & \sigma_{\delta} 
\end{pmatrix}
\end{align*}
where $(\sigma_{\lambda}, \sigma_{f}, \sigma_{\phi}, \sigma_{\delta}) \sim Exponential(1)$ and $\Omega \sim LKJ(2)$ \citep{Lewandowski2009}. Our main results are not sensitive to the specification of priors we use.
\end{small}}

We estimate the model with preference biased prior beliefs. In the statistical model, the probability that participant $i$ chooses alternative $j$ is:
\begin{align*}
\label{eq:stat_model}
Pr(Y_{i} = j) = \frac{e^{\lambda_{it} \nu_{ij} } (s_{ij})^{f_{itc}} }{\sum_k e^{\lambda_{it} \nu_{ik} } (s_{ik})^{f_{itc}} }
\end{align*}

The model parameter $\lambda_i$ reflects the effect of the estimated intrinsic utility $\nu_{ij}$ on the individual's choice. The exponential form implies that the choice probability is a function of the difference in intrinsic utility of both alternatives that we measure by $\Delta$. For convenience we use $\nu_{ij} := \frac{\Delta}{2}$ if $j$ is predicted to be the preferred alternative, and $\nu_{ij} := -\frac{\Delta}{2}$ otherwise. The exponential form of the intrinsic preference component assures that different distributions of $\Delta$ over the two choice alternatives will give the same results. 

The model parameter $f_{itc}$ reflects the effect of social learning based on $s_{ij}$, the frequency of alternative $j$ expected by the participant. The expected frequency of alternative $j$ is: 
\begin{align*}
s_{ij} = \frac{2\phi_{it}\frac{e^{ \delta_{it} \nu_{ij}}}{\sum_k e^{\delta_{it} \nu_{ik}}} + n_{ij}}{2 \phi_{it} + 5}
\end{align*}
The parameter $\phi_{it}$ is the strength of the prior and can be interpreted as the number of past observations of each alternative. The parameter $\delta$ reflects the preference bias of the prior belief. As before, $n_{ij}$ indicates the number of observed uninfluenced participants that chooses alternative $j$.

\subsection*{Simulations on social norms and preferences}
We run simulations to assess whether the behavior of the majority contains information about the average preferences in a population with heterogeneous preferences. We simulate the choices of $N = 50$ individuals over 50 periods between a commonly preferred alternative and another alternative. For a specific combinations of model parameters, we repeat the simulation 50 times, and report the average share of a target alternative across the 50 simulations. 

We assume a standard normal distribution of preferences. Across the simulations we vary the mean preferences in steps of 0.1 ranging from 0 (no common preference) to 4 (strong common preferences). 

For the other model parameters, we assume no heterogeneity. We vary the social learning parameter $f$ from -2 (anticonformity) to 2 (conformity) in steps of 0.1. As possible strength of the prior beliefs we use values of 1, 5, and 20. The preference-bias of prior beliefs is either 0, 0.5 or 1. To also assess the influence of an extreme minority, we use a second set of simulation, for which we assume that every fifth individual has preferences of -4, with no heterogeneity.

\begin{figure}[htbp]
\caption{Simulation}
\label{fig:simulation}
\begin{center}
\includegraphics[scale=0.65]{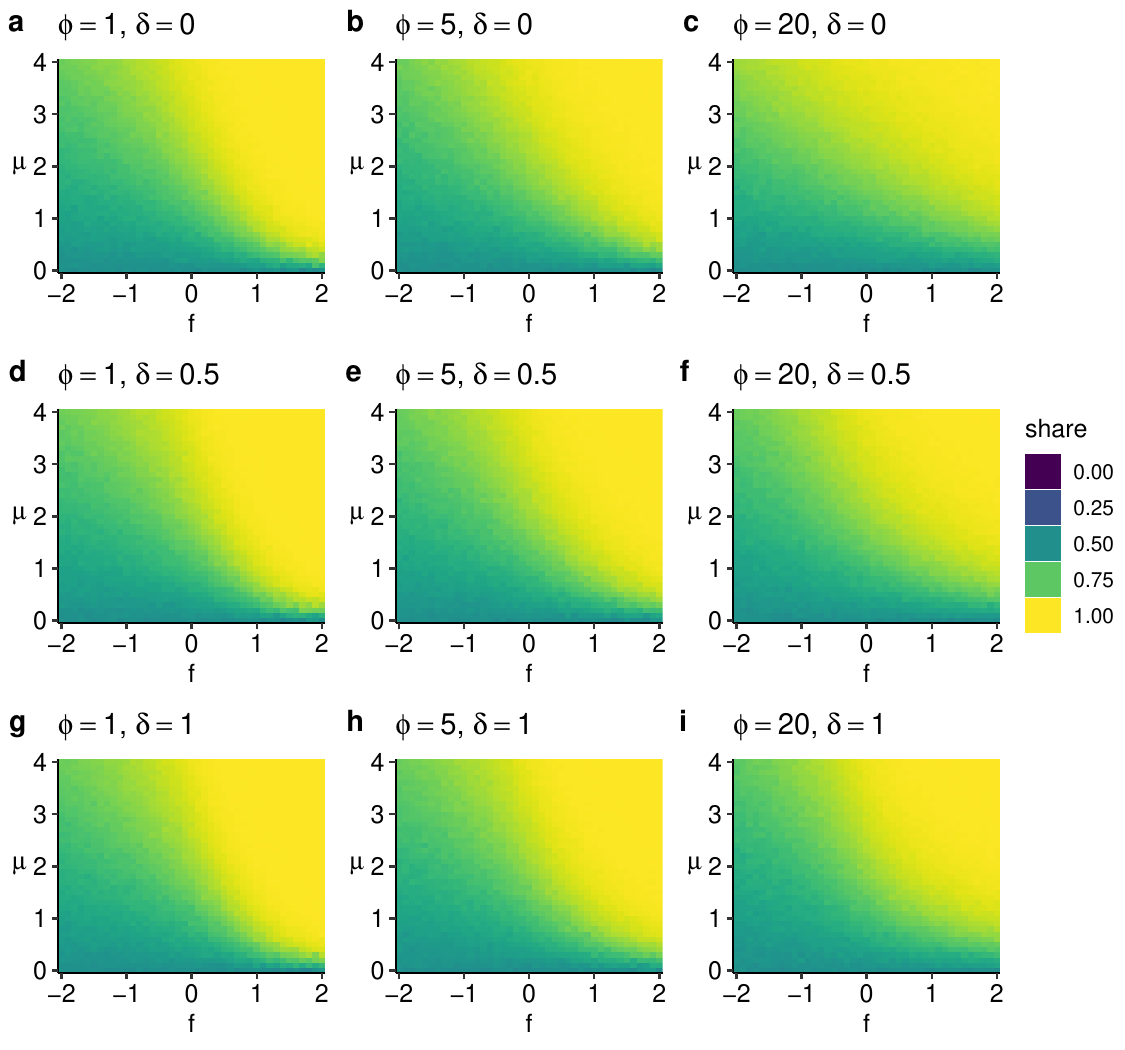}
\end{center}
\end{figure}

\begin{figure}[htbp]
\caption{Simulation with minority}
\label{fig:simulation_minority}
\begin{center}
\includegraphics[scale=0.65]{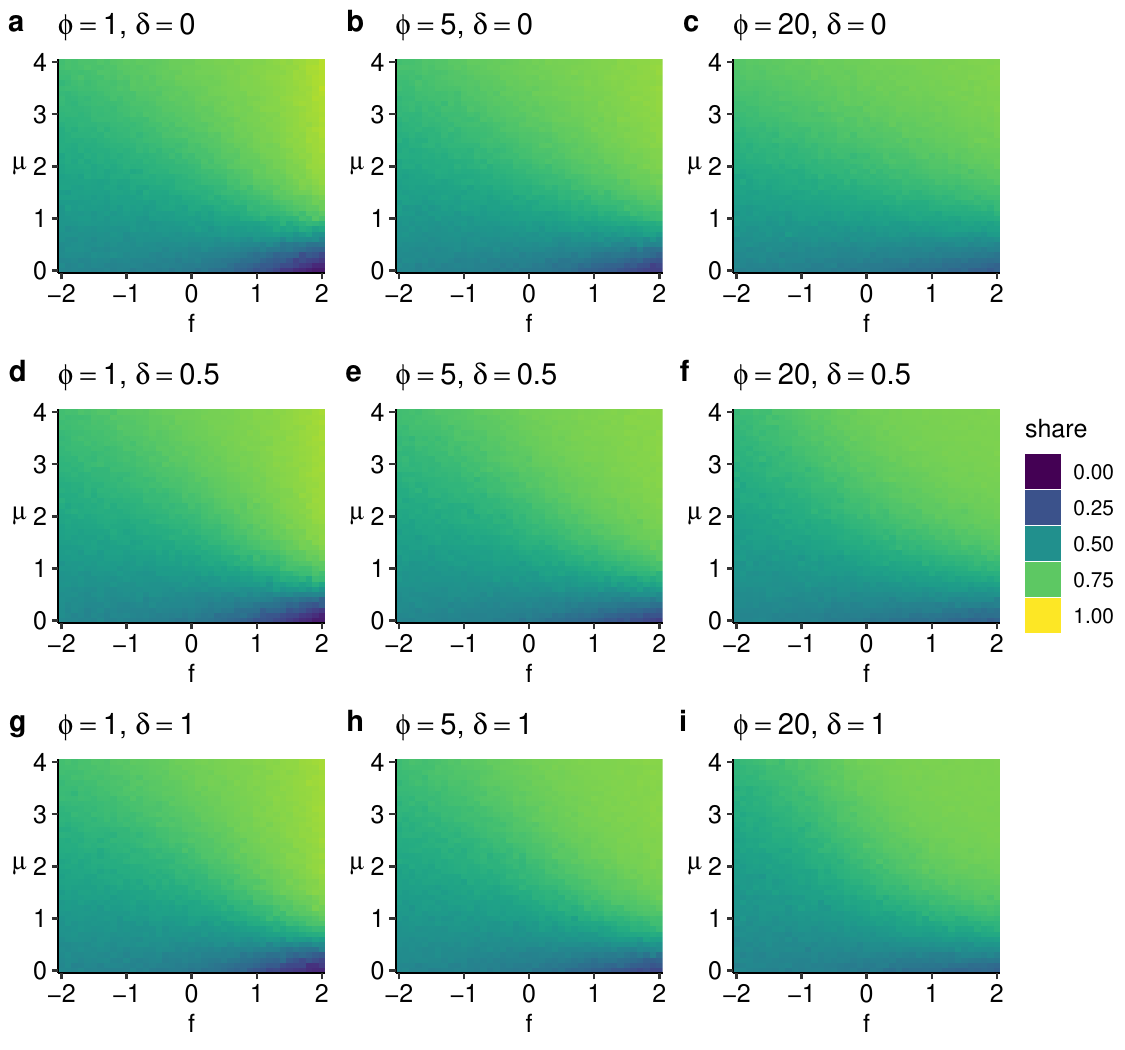}
\end{center}
\end{figure}

Figures \ref{fig:simulation} and \ref{fig:simulation_minority} show that the average share of the commonly preferred alternative in the last period of the 50 independent simulation runs. The color coding of the graphs shows averages above one-half in green and yellow colors. These colors are usually more frequent, which suggests that the descriptive social norm usually contains information about the average preferences of the population. An exception to this rule applies with an extreme minority and weak majority preferences. In these cases, conformist social learning induces convergence on the alternative preferred by the minority.

\subsection*{Study materials and instructions}
\label{app:study materials}
\begin{figure}[htbp]
\caption{Decision screen with slider}
\label{fig:paintings_slider}
\begin{center}
\includegraphics[scale=0.5]{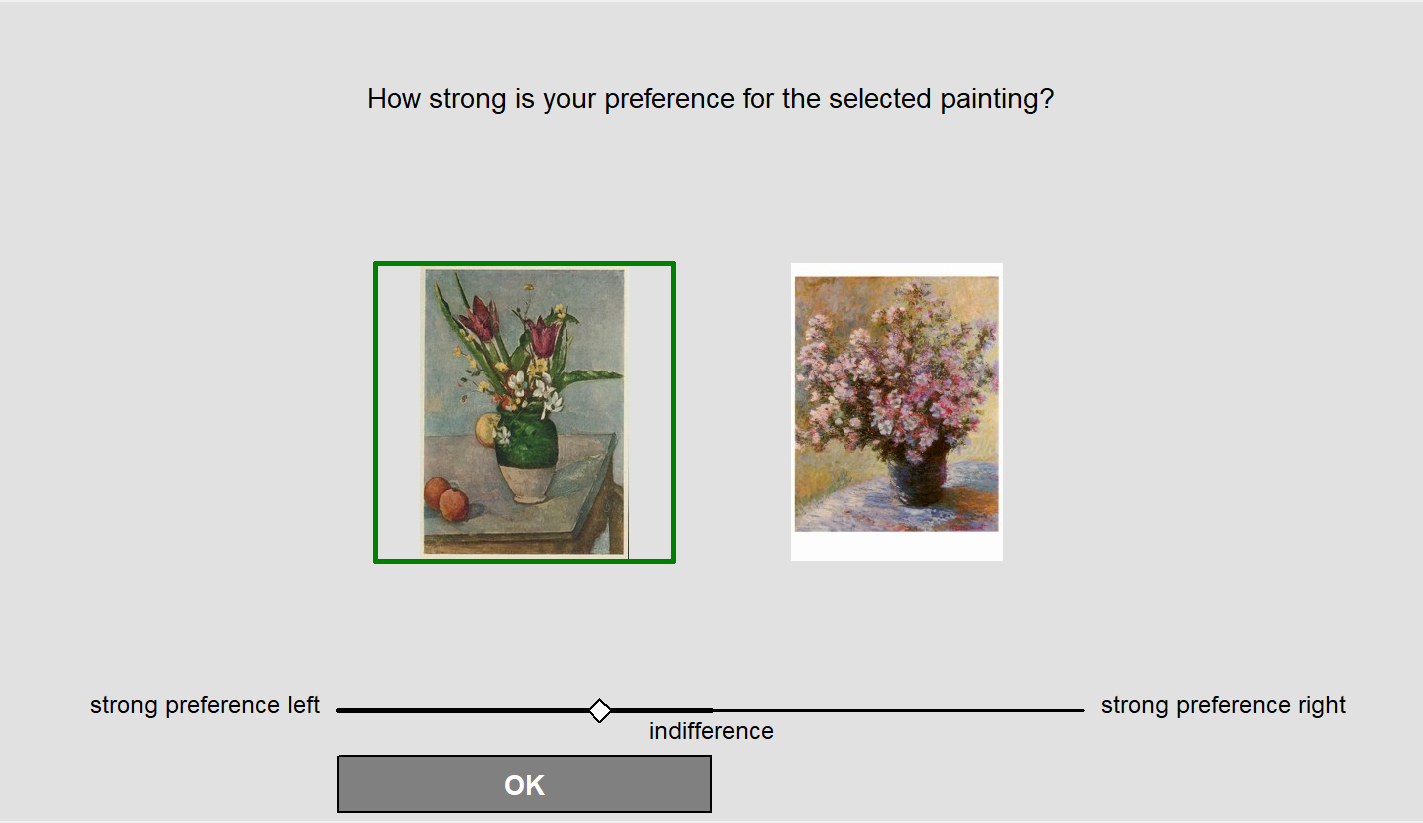}
\end{center}
\end{figure}

\begin{figure}[htbp]
\caption{Evaluation screen}
\label{fig:paintings_slider}
\begin{center}
\includegraphics[scale=0.5]{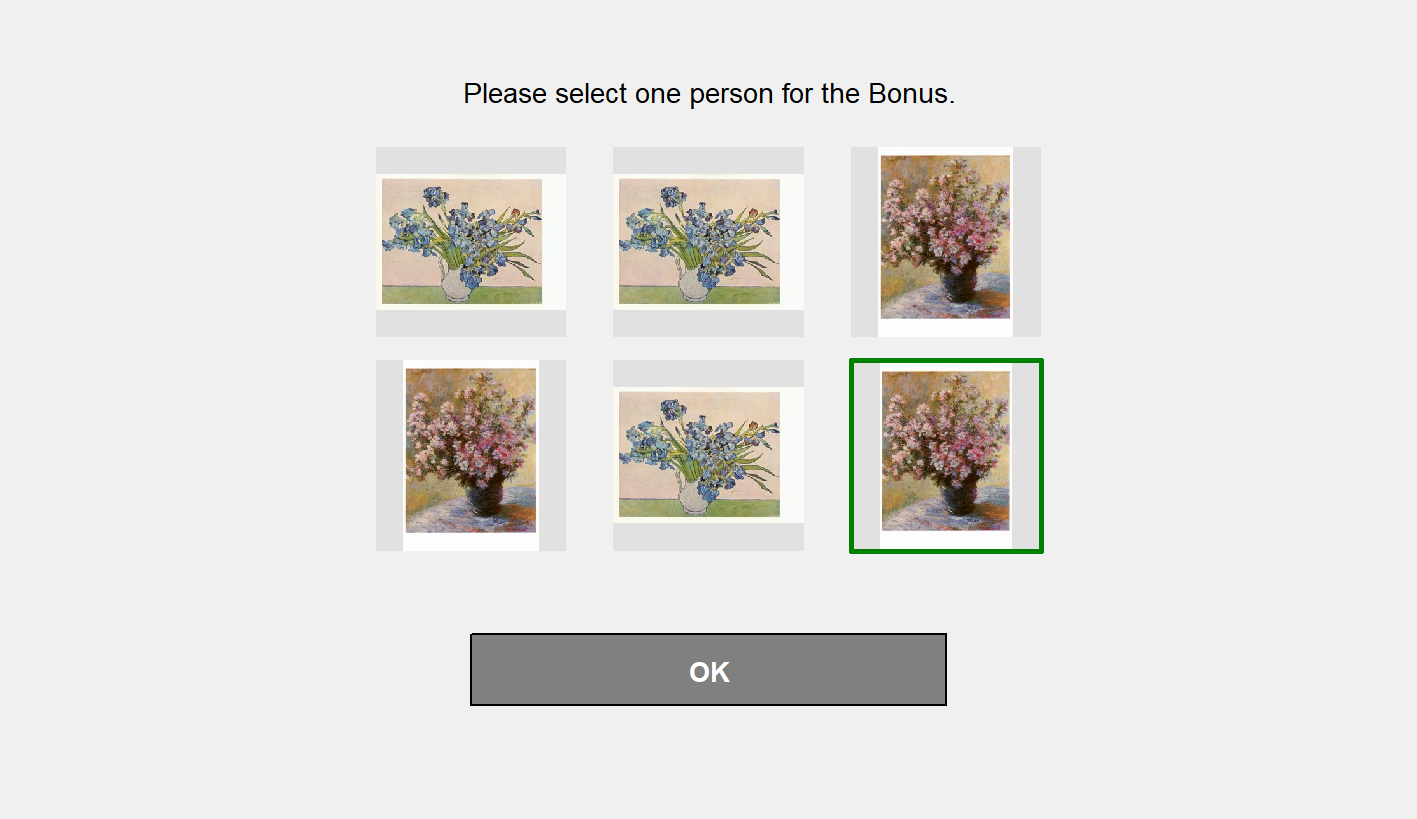}
\end{center}
\end{figure}

\begin{sidewaystable}[htbp]
	\caption{List of paintings}
	\label{tab:paintings}
	\setlength{\tabcolsep}{5pt}
	\renewcommand\arraystretch{0.8}
	\setlength{\tabcolsep}{3pt}
	\centering 	
	\begin{scriptsize}
		\begin{tabular}{llllllll}
					\hline\hline \\[-2ex]
			{\footnotesize Set} & {\footnotesize Theme} & {\footnotesize Painting 1}   &  {\footnotesize Painting 2}  &    \\
						\hline
1  & Marc Chagall 	& Bride and Groom &  The couple \\
2  & Lyonel Feininger& The Grain Tower at Treptow on the Rega & Village Pond of Gelmeroda \\
3  & Claude Monet	& The Artist's Garden in Giverny& 	The artist's Garden in Giverney \\
4  & Wassily Kandinsky & 	Improvisation 26 & Improvisation 28 (2nd version) \\
5  & August Macke	& Garden Restaurant& Large Bright Walk \\
6  & Franz Marc 	& Yellow Cow & The Dog in Front of the World \\
7  & Caspar D. Friedrich  & The Churchyard Gate & Hutten's Grave (Ruin of a Church Choir )  \\
8  & Houses 		& Egon Schiele,	House with drying Laundry & Albrecht Duerer, The Castle of Trient \\
9  & Work			& Paul C\'{e}zanne, The Mowers & Vincent van Gogh, Field with Farmer and Mill \\
10 & Women 			& August Macke, Portrait with Apples  & Pablo Picasso, The Absinthe-Drinker \\
11 & Trees 			& Paul C\'{e}zanne,	The Chestnut Trees at Jas de Bouffan & Rudolph von Alt, Landscape in the Prater in Vienna \\
12 & Hands 			& Albrecht Duerer, The Hands of Jesus Christ & Pablo Picasso, Crossed Hands \\
13 & Ships			& Egon Schiele, Fishing Boats in Trieste & 	Berthe Morisot, The Harbour of Nice \\
14 & Flowers 		&Paul C\'{e}zanne, Flowers in a Vase and Fruit & Vincent van Gogh,	Bouquet of Irises \\
15 & Bridges 		& Claude Monet, Bridge over the Seine near Argenteuil & Vincent van Gogh, The Bridges of Asni\`{e}res \\
\hline
{\footnotesize Set} & {\footnotesize Theme} & {\footnotesize Painting 3}   &  {\footnotesize Painting 4 } \\
\hline
1  & Marc Chagall 	& The Newly-Married of the Eiffel-Tower & Lovers\\
2  & Lyonel Feininger &  Gelmeroda IX & The Church of Halle\\
3  & Claude Monet	& The artist's Garden in V\'{e}theuil & Resting under the Lilac\\
4  & Wassily Kandinsky  & Improvisation 34 (Orient II) & Improvisation Gorge\\
5  & August Macke	 & Girls under Trees & Sunny Path\\
6  & Franz Marc 	 & 	The Tiger & 	Fox (Blue Black Fox) \\
7  & Caspar D. Friedrich  &	Ruins of the Monastery Eldena & The Graveyard Door (The Churchyard) \\
8  & Houses 		& Rudolph von Alt, The "Goldene Dachl" in Innsbruck & Carl Spitzweg, A Hypochondriac \\
9  & Work			& Carl Spitzweg, The Walk of the Boarding School & Pond at the Forest\\
10 & Women 			& Egon Schiele, Peasants-Girl & Gustav Klimt, Johanna Staude\\
11 & Trees 			& Vincent van Gogh, Road with Cypress and Star & Alfred Sisley, The Path to the Old Ferry at By\\
12 & Hands 			& Egon Schiele, Clasped Hands & August Rodin, The Cathedral - Hands \\
13 & Ships			& Redon Odilon, The Mystical Boat & Raoul Dufy,	The old Harbour of Marseille\\
14 & Flowers 		& Claude Monet, Vase of Flowers & Pierre-Auguste Renoir, Bouquet of Chrysanthemums\\
15 & Bridges 	    & Alfred Sisley, Bridge near Hampton Court & 	William Turner, Old Welsh Bridge, Shrewsbury\\
		\hline\hline \\[-2ex]
		\end{tabular}
	\end{scriptsize}	
	Before each session, we selected 10 sets and for each set three of the paintings listed based on availability. Postcards of the paintings were ordered from Kunstverlag Reisser, Braunschweigstrasse 12, 1130 Vienna, Austria. Names are translations from German and taken from http://www.reisser-kunstpostkarten.de.
\end{sidewaystable}
\setstretch{1}

\begin{sidewaystable}[htbp]
	\caption{Question sets 1-9}
	\label{tab:questions1}
	\renewcommand\arraystretch{0.8}
	\setlength{\tabcolsep}{1pt}
	\centering 	
	\begin{scriptsize}
		\begin{tabular}{llllllllllll}
					\hline\hline \\[-2ex]
			Set & Question & Option 1   &  Option 2  &  Option 3 & Answer 1 & Answer 2 & Answer 3 \\   [1ex] 
			\hline  
    1 & Which country is larger (2015, m\^2)? & Canada  & USA   & China &  9984k & 9826k & 9596k \\
    1 & Which country is larger (2015, m\^2)? & Portugal & Czech Republic & Austria &  92090 & 78867 & 83871 \\
    1 & Which country is larger (2015, m\^2)? & Estonia & Denmark & Netherlands & 45228 & 43094 & 41543 \\
    1 & Which country is larger (2015, m\^2)? & Lithuania & Croatia & Latvia &  65300 & 56594 & 64589 \\
    1 & Which country is larger (2015, m\^2)? & Sudan & Indonesia & Mexico &  1,861k & 1,904k & 1,964k \\
    2 & Which country has more inhabitants (2014)? & France & Italy & UK    &  65,835k & 60,782k & 64,351k \\
    2 & Which country has more inhabitants (2014)? & Spain & Ukraine & Poland &  46,512k & 45,245k & 38,017k \\
    2 & Which country has more inhabitants (2014)? & Greece & Belgium & Czech Republic &  10,926k & 11,203k & 10,512k \\
    2 & Which country has more inhabitants (2014)? & Austria & Switzerland & Bulgaria &  8,506k & 8,139k & 7,245k \\
    2 & Which country has more inhabitants (2014)? & Malta & Luxemburg & Iceland &  425k & 549k & 325k \\
    3 & Which company had more employees (2014)? & Bosch & Daimler & Metro &  290,183 & 279,972 & 24,9150 \\
    3 & Which company had more employees (2014)? & Bayer & ThyssenKrupp & Continental  &  118,900 & 160,745 & 189,168 \\
    3 & Which company had more employees (2014)? & Lufthansa & BASF  & BMW   &  118,781 & 113,292 & 116,324 \\
    3 & Which company had more employees (2014)? & RWE   & E.ON  & MAN   &  59,784 & 58,503 & 55,903 \\
    3 & Which company had more employees (2014)? & Bertelsmann  & SAP   & TUI   &  112,037 & 74,406 & 77,309 \\
    4 & Who was born earlier?  & Konrad Adenauer & F.D. Roosevelt & Theodor Heuss &  1876 & 1882 & 1884 \\
    4 & Who was born earlier?  & Willy Brandt & John F. Kennedy & Walter Scheel &  1913 & 1917 & 1919 \\
    4 & Who was born earlier?  & Helmut Schmidt & Richard Nixon & R. Weizsaecker &  1918 & 1913 & 1920 \\
    4 & Who was born earlier?  & Horst Koehler & Gerhard Schroeder & Bill Clinton &    1943 & 1944 & 1946 \\
    5 & Which harbor is bigger (2014, TEU)? & Shanghai & Hong Kong & Singapore &  35.3  & 22.30 & 33.9 \\
    5 & Which harbor is bigger (2014, TEU)? & Hamburg  & Antwerp & Los Angeles &  9.7   & 9     & 8.3 \\
    5 & Which harbor is bigger (2014, TEU)? & Guangzhou & Dubai & Rotterdam &  16.2  & 15.2  & 12.3 \\
    6 & Which airline had more passengers? & United Airlines & American Airlines  & Ryanair & 90,440k & 87,830k & 86,370k \\
    6 & Which airline had more passengers? & Lufthansa & Easyjet & Air China &  59,850k & 62,310k & 54,580k \\
    6 & Which airline had more passengers? & Air Berlin & Brithish Airlines & Air France &  29,910k & 41,160k & 45,410k \\
    6 & Which airline had more passengers? & KLM   & Aeroflot & SAS  & 27,740k & 23,600k  & 27,390k \\
    7 & Which country discharges more CO2 (2010, pp)? & Germany & Netherlands & Austria & 12.3  & 10.1  & 12.1 \\
    7 & Which country discharges more CO2 (2010, pp)? & Poland & Slovakia & Hungary &  7.7   & 7.8   & 7.3 \\
    7 & Which country discharges more CO2 (2010, pp)? & Lithuania & Latvia & Estonia &  5.9   & 6.5   & 13.5 \\
    7 & Which country discharges more CO2 (2010, pp)? & France & Portugal & Spain &  9     & 6.9   & 8.5 \\
    7 & Which country discharges more CO2 (2010, pp)? & Finland & Norway & Sweden  &  18.7  & 10.1  & 9.3 \\
    8 & Which country has more inequality (2012, GINI)? & France & Belgium & Austria &  33.1  & 27.6  & 30.5 \\
    8 & Which country has more inequality (2012, GINI)? & Norway & Finland & Sweden &  25.9  & 27.1  & 27.3 \\
    8 & Which country has more inequality (2012, GINI)? & Bolivia & Ecuador & Peru  &  46.7  & 46.6  & 45.1 \\
    8 & Which country has more inequality (2012, GINI)? & Costa Rica & Brazil & Argentina &  48.6  & 52.7  & 42.5 \\
    8 & Which country has more inequality (2012, GINI)? & Thailand & Laos  & Vietnam &  39.3  & 37.9  & 38.7 \\
    9 & Which soccer club is worth more (2016)? & Manchester City & FC Chelsea & M. United &  501.75 & 490   & 411.25 \\
    9 & Which soccer club is worth more (2016)? & AS Rom & FC Valencia  & SSC Neapel &  250.7 & 282   & 284 \\
    9 & Which soccer club is worth more (2016)? & Bayer 04 Leverkusen  & VfL Wolfsburg & FC Schalke 04 &  211.1 & 183.1 & 199.8 \\
    9 & Which soccer club is worth more (2016)? & Zenit St.Petersburg & AC Mailand & FC Sevilla &  198.6 & 188.1 & 186.2 \\
 \hline\hline \\[-2ex]
		\end{tabular}
	\end{scriptsize}	
\end{sidewaystable}
\setstretch{1}

\begin{sidewaystable}[htbp]
	\caption{Question sets 10-19}
	\label{tab:questions2}
	\setlength{\tabcolsep}{5pt}
	\renewcommand\arraystretch{0.8}
	\setlength{\tabcolsep}{1pt}
	\centering 	
	\begin{scriptsize}
		\begin{tabular}{llllllllllll}
					\hline\hline \\[-2ex]
			Set & Question & Option 1   &  Option 2  &  Option 3 & Answer 1 & Answer 2 & Answer 3 \\   [1ex] 
			\hline  
    10 & Which country won more medals (2014 Olympics?   & Netherlands & France & Germany &  24    & 15    & 19 \\
    10 & Which country won more medals (2014 Olympics?   & Switzerland & Sweden & Austria &  11    & 15    & 17 \\
    10 & Which country won more medals (2014 Olympics?   & Canada  & Norway & USA   &  26    & 28 \\
    10 & Which country won more medals (2014 Olympics?   & Finland & UK    & Ukraine     & 5     & 4     & 2 \\
    10 & Which country won more medals (2014 Olympics?   & Belarus & Kazakhstan & Australia      & 6     & 1     & 3 \\
    11 & Which airport has more passengers (2014)? & Atlanta Int & L Heathrow & Dubai Int & 96,178k & 73,408k & 70,475k \\
    11 & Which airport has more passengers (2014)? & Singapore Changi & Kuala Lumpur & Shanghai Int &54,093,000 & 48,930k & 51,687k\\
    11 & Which airport has more passengers (2014)? & Charles de Gaulles & Frankfurt & A Schiphol & 63,813k & 59,566k & 54,978k \\
    11 & Which airport has more passengers (2014)? & Madrid Barajas & SP-Guarulhos & Miami Int & 41,822k & 39,765k & 40,941k \\
    12 & Who sold more records in Germany? & The Beatles & Michael Jackson & Madonna & 7,600k & 11,275k & 12,300k \\
    12 & Who sold more records in Germany? & ACDC  & ABBA  & R. Williams & 10,475k & 10,800k & 9,275k \\
    12 & Who sold more records in Germany? & Helene Fischer & Pur   & Die Aerzte & 9,150k & 9,425k & 7,850k \\
    12 & Who sold more records in Germany? & Britney Spears & Bon Jovi & Xavier Naidoo & 5,050k & 5,150k & 5,525k \\
    13 & In which language is the letter "a" more frequent?  & German & English & French & 6.51  & 8.167 & 7.636 \\
    13 & In which language is the letter "a" more frequent?  & Spanish & Italian & Swedish  & 12.53 & 11.740 & 9.300 \\
    13 & In which language is the letter "a" more frequent?  & German & English  & French & 17.4  & 12.702 & 14.715 \\
    13 & In which language is the letter "a" more frequent?  & Spanish & Italian & Swedish  & 13.68 & 11.790 & 9.900 \\
    13 & In which language is the letter "a" more frequent?  & Spanish & Italian & Swedish  & 6.71  & 6.880 & 8.800 \\
    14 & Which initial letter is more common in German? & E     & I     & W     & 7.8   & 7.1   & 6.8 \\
    14 & Which initial letter is more common in German? & H     & I     & O     & 7.232 & 6.286 & 6.264 \\
    14 & Which initial letter is more common in German? & C     & D     & F     & 3.511 & 2.670 & 3.779 \\
    14 & Which initial letter is more common in German? & J     & K     & V     & 0.597 & 0.590 & 0.649 \\
    15 & Which country has more prisoners (2016, per 100k)? & USA   & Cuba  & Seychelles & 698   & 510   & 799 \\
    15 & Which country  has more prisoners (2016, per 100k)? & Thailand & Russia & Ruanda & 468   & 447   & 434 \\
    15 & Which country has more prisoners (2015, absolute)? & Berlin & Saxony & Rhineland & 3806  & 3385  & 3102 \\
    15 & Which country has more prisoners (2015, absolute)? & Saxony-Anhalt & Thuringia & Hamburg & 1670  & 1600  & 1559 \\
    15 & Which country has more prisoners (2015, absolute)? & Schleswig-Holstein & Mecklenburg & Brandenburg & 1162  & 1057  & 1324 \\
    16 & Which food has more calories (per 100g)? & Paprika Yellow & Paprika Red & Paprika Green &  28    & 33    & 20 \\
    16 & Which food has more calories (per 100g)? & Rhubarb & Radicchio & Peperoni &  14    & 13    & 20 \\
    16 & Which food has more calories (per 100g)? & Zucchini & Spinach & Pak Choi & 18    & 15    & 16 \\
    16 & Which food has more calories (per 100g)? & Leek  & Broccoli & Red cabbage & 24    & 26    & 22 \\
    16 & Which food has more calories (per 100g)? & Wild garlic & Eggplant & Artichoke &  19    & 17    & 22 \\
    17 & Which country is older? & Albania & Finland & Hungary & 1912  & 1917  & 1918 \\
    17 & Which country is older? & New Zealand & Norway & Panama & 1907  & 1905  & 1903 \\
    17 & Which country is older? & Ghana & Niger & Togo  & 1957  & 1958  & 1960 \\
    17 & Which country is older? & Tanzania & Ruanda & Mali  & 1964  & 1962  & 1960 \\
    17 & Which country is older? & Brazil & Uruguay & Costa Rica & 1822  & 1825  & 1821 \\
    18 & Which country has more internet users (2015)? & Austria & Germany & UK    & 83.1  & 88.4  & 91.6 \\
    18 & Which country has more internet users (2015)? & Luxemburg & Netherlands & Denmark & 94.7  & 95.5  & 96 \\
    18 & Which country has more internet users (2015)? & Portugal & Italy & Greece &  67.9  & 62    & 63.2 \\
    18 & Which country has more internet users (2015)? & Myanmar & Laos  & Nepal &  12.6  & 14.3  & 18.1 \\
    18 & Which country has more internet users (2015)? & Jamaica & Peru  & Panama & 53.6  & 52.6  & 52 \\
    19 & Which country has more analphabets (relative)? & Guinea & Niger & Burkina Faso & 74.7  & 84.5  & 71.3 \\
    19 & Which country has more analphabets (relative)? & Mali  & Chad  & Ethiopia &  66.4  & 62.7  & 61 \\
    19 & Which country has more analphabets (relative)? & Liberia & Haiti & Sierra Leone &  57.1  & 51.3  & 55.5 \\
    19 & Which country has more analphabets (relative)? & Pakistan  & Bhutan & Senegal &  45.3  & 47.2  & 47.9 \\
    19 & Which country has more analphabets (relative)? & Nigeria & Mozambique & Gambia & 48.9  & 49.4  & 48 \\
    		\hline\hline \\[-2ex]
		\end{tabular}
	\end{scriptsize}	
\end{sidewaystable}

\newpage

\subsection*{Instructions of the online experiment}
\vspace{0.4cm}
\begin{center}
{\Large Instructions}
\end{center}

We ask you to keep quiet at your computer workstation and not to communicate with others during the experiment. Anyone who intentionally violates this rule is requested to leave the experiment without payment.
If you have any questions, please contact us and wait until an experimenter comes to you.
Income is calculated in points. At the end of the experiment, the total amount of points you have earned will be converted into euros at the following rate: 

\begin{center}
1 point = 1 Euro
\end{center}

You will receive your total income in cash at the end of the experiment. 

Please read the instructions carefully. When all the instructions have been read, answer some comprehension questions. Then make your decisions in the experiment. Your decisions will be treated anonymously.\smallskip\\

\noindent {\large General procedure }\vskip5pt

This experiment consists of two parts, each consisting of three stages. In each step you make several decisions. \smallskip\\

You have a certain number of seconds for each stage. The remaining time is always displayed on the top right of the screen. The time for each stage is such that it should be sufficient to process all decisions. When the time is up, the experiment automatically continues. The remaining decisions are then randomly generated by the computer. If all participants have completed all the decisions of a stage before the end of the specified time, the experiment continues immediately.\smallskip\\
For each decision you work on within the given time, they will be credited a certain amount of Experimental Currency Units. Your total income is the sum of the EWE from all parts in Euro. \smallskip\\

In this experiment, we share with other participants some of your decisions. Since the other participants will not receive any other information about you, your decisions will remain anonymous. You will also be shown decisions from other participants. If other participants have not processed a decision in the given time, you will be informed that the displayed decisions are randomly generated decisions. \smallskip\\

\noindent {\large{Overview Part 1}}\vskip5pt

Below are the instructions for Part 1. You will receive the instructions for Part 2 when Part 1 is completed. Your decisions in Part 1 do not affect your income in Part 2.

\begin{center}
{\Large Which postcard do you choose?}
\end{center}

For the first part, you will be randomly divided into groups of 6 people. In this part, you will each decide between two art postcards. To do this, you will be shown the designs of the two cards on the screen and you will choose which of the two designs you would prefer to have. These choices are made for multiple pairs of postcards. The other members of your group also make such decisions.

At the end of the experiment, a decision situation will be randomly selected for your group. You will then be sent your preferred design as a postcard per mail. 

This part consists of 3 stages that you will work on one after the other. The stages are described in more detail below.

Stage 1

On the screen you will see two postcard motifs A and B. You decide which of the two motifs you prefer. To do so, you simply click the corresponding motif and confirm your choice later.

After each decision, we ask you to indicate how strong your preference is for the motif you selected. To do this, a slider will appear below the paintings after your decision. Please try this out by clicking on one of the motifs below.

You make these choices 50 different combinations of postcards. For each decision processed within the given time, you are credited 0.5 EWE.

Stage 2

Also in this stage, you decide between two postcard motifs and indicate how strong your preference is for the chosen motif.

Before each decision, you will be informed about the decisions of the five other members of your group in the same choice. However, the other participants have already gone through the decision situation in stage 1. That is, they had no information about the decisions of other participants at the time of their decision. The paintings chosen by the others are displayed in random order on your screen. An example of the display can be found below.

You make this decision one by one for 10 pairs of postcards. You will be credited 0.5 EWE for each edited decision of the second part within the given time.

Stage 3 

In Stage 3, the 6 selected paintings of each group are transmitted to a member of another group for evaluation. Based on the selected paintings, the member of the other group selects one person in the group to receive a Bonus of 10 points. You will also judge the choices of participants of another group.

All evaluators work on several decision situations of the other group. You do not know which decisions were made in stage 1 and which in stage 2. Also, the positions where you see the paintings of the six group members are randomized for each decision situation. 

In addition, all evaluators process several more randomly generated decision situations. They do not know which decisions are real and which are randomly generated decision situations. At the end of the first part, a real decision situation is drawn for each group. The person selected in this decision situation will then receive the bonus. Below is an example of the judgment situation that results from the example in Stage 2. One member of your group has chosen  painting A and five have chosen painting B. Your decision appears at the bottom right next to center. You can now take the role of the member of the other group and click on one of the six paintings. The painting of the selected person will be highlighted by a green frame.

control questions 

Please use the control questions to check that you have understood the instructions correctly. Your answers will have no effect on the experiment. 

Q1: How many blocks does this experiment have?
Q2: How many parts does each block have?
Q3: The decisions of how many other participants do you see in part 2?
Q4: In which part did the other participants make their decisions?
Q5: What happens to the unprocessed decisions of a person when the time for a stage has expired? These will be...? (randomly generated, repeated later)
Q6: What applies always when a decision was randomly generated? (I will be informed, I will not be informed)

conclusion

Finally, you will receive an overview of your income from the experiment. We will transfer the money to the account you specified.

In addition, you will receive the design you chose in the randomly selected decision situation as a postcard. The postcard will be sent to you by mail in the next few days.

\end{document}